\documentclass[trackchanges]{aastex701}

\usepackage{enumitem}

\begin{document}

\title{A Critical Evaluation of the Physical Nature of the Little Red Dots}

\newcommand{\mbh}{M_{\rm BH}}
\newcommand{\msun}{M_\odot}
\newcommand{\rsun}{R_\odot}
\newcommand{\zsun}{Z_\odot}
\newcommand{\lsun}{L_\odot}
\newcommand{\gcc}{{\rm g~cm}^{-3}}
\newcommand{\cc}{{\rm cm}^{-3}}
\newcommand{\mdot}{\dot m}
\newcommand{\msunyr}{M_\odot~{\rm yr}^{-1}}
\newcommand{\mdote}{\dot{\rm M}_{\rm Edd}}
\newcommand{\kpc}{{\rm kpc}}
\newcommand{\mpc}{{\rm Mpc}}
\newcommand{\gpc}{{\rm Gpc}}
\newcommand{\pc}{{\rm pc}}
\newcommand{\mum}{\mu {\rm m}}
\newcommand{\kms}{{\rm km~s}^{-1}}
\newcommand{\ergs}{{\rm erg~s}^{-1}}
\newcommand{\ledd}{{\rm L}_{\rm Edd}}

\newcommand{\Muv}{M_{\rm UV}}
\newcommand{\SFR}{{\rm SFR}}
\newcommand{\yr}{\rm yr}
\newcommand{\Hmol}{\rm H_2}
\newcommand{\K}{{\rm K}}
\newcommand{\beq}{\begin{equation}}
\newcommand{\eeq}{\end{equation}}
\newcommand{\muJy}{\mu{\rm Jy}}
\newcommand{\D}{{\rm d}}

\newcommand{\red}[1]{\textcolor{red}{ #1}}
\newcommand{\blue}[1]{\textcolor{blue}{ #1}}
\newcommand{\cyan}[1]{\textcolor{cyan}{ #1}}

\author[orcid=0000-0000-0000-0001,sname='Inayoshi']{Kohei Inayoshi}
\altaffiliation{}
\affiliation{Kavli Institute for Astronomy and Astrophysics, Peking University, Beijing 100871, China}
\email[show]{inayoshi@pku.edu.cn}  

\author[0000-0001-9840-4959,sname='Ho']{Luis C. Ho}
\altaffiliation{}
\affiliation{Kavli Institute for Astronomy and Astrophysics, Peking University, Beijing 100871, China}
\affiliation{Department of Astronomy, School of Physics, Peking University, Beijing 100871, China}
\email[]{}

\collaboration{all}{}

\begin{abstract}
Little Red Dots (LRDs) are a newly identified class of active galactic nuclei (AGNs) uncovered by JWST deep surveys. 
Their enigmatic properties challenge the canonical AGN paradigm and have stimulated numerous ideas on 
the early formation and growth mechanisms of massive black holes (BHs).
In this review, we summarize how early BHs can shape the characteristic features of LRDs, 
how their nuclear environments differ from those of normal AGNs,
and how future observations can distinguish between competing scenarios (AGN or stellar origins).
Our main conclusions are as follows:\vspace{3pt}

\begin{itemize}[labelsep=-1em, leftmargin=5em, align=parleft, topsep=0em, parsep=-2pt]

\item LRDs show broad-line emission consistent with mass accretion onto BHs with $M_{\rm BH}\simeq 10^{6}-10^{7}~\msun$,
suggesting that AGN activity is a plausible origin of their dominant red optical emission.

\item Stellar components can reproduce the continuum energetics through dusty star formation.
However, the required stellar mass would be too large to remain consistent with other key LRD properties.
Therefore, a purely stellar origin is unlikely to be the dominant power source, although star formation may still 
contribute to the UV emission.

\item  The coexistence of broad-line emission with Balmer absorption and break features on LRD spectra,
neither of which can be explained by evolved stellar populations, suggests that nuclear BHs
are enshrouded by dense gas with a high covering fraction.

\item Gas-enshrouded AGNs can produce red optical spectra without requiring dust reddening
through a combination of gas attenuation and thermal self-emission with an effective temperature 
of $T_{\rm eff}\simeq 5000~\K$, which also accounts for the flat infrared continuum.
This scenario therefore offers a compelling explanation, although alternative interpretations are discussed in this review.

\item From the spectral features and redshift evolution,
LRDs are likely a transient phase in early BH growth, possibly the first accretion episodes of newborn BHs.

\item Testing models for LRD spectra and origins through exploring time variability, 
sources of ionizing radiation, post-LRD objects, and low-redshift LRD analogs is particularly promising.

\end{itemize}
\end{abstract}


%\tableofcontents

\section{Introduction}

The James Webb Space Telescope (JWST) has opened an unprecedented window into high-redshift extragalactic astronomy.
Thanks to its sensitivity and spatial/spectral resolution in the infrared (IR) bands, redshifted light from active galactic nuclei (AGNs) powered 
by accreting massive black holes (BHs) can be observed in detail, offering new insights into the nature and evolution of 
early BHs and their host galaxies.
One of the most remarkable discoveries by JWST is a new population of AGNs known as little red dots (LRDs), 
compact sources with red optical continua and broad-line emission signatures. 
The first spectroscopically confirmed LRD, CEERS~746 at $z = 5.624$, was reported by \citet{Kocevski_2023}.
Following this discovery, a slitless spectroscopic survey with JWST/NIRCam identified approximately 20 similar red sources 
exhibiting broad H$\alpha$ emission, earning them the nickname ``Little Red Dots" \citep{Matthee_2024}. 
Independent studies have reported additional LRDs from multiple JWST programs, including CEERS \citep{Harikane_2023_agn}, 
JADES \citep{Maiolino_2024_JADES}, UNCOVER \citep{Greene_2024}, and ASPIRE \citep{Lin_2024} and 
confirmed that LRDs are a common and robust phenomenon in the early universe.

The discovery of LRDs has sparked considerable interest in both observational and theoretical communities, 
as LRDs may represent a previously unrecognized, early phase of BH accretion activity and thus potentially
reshape our understanding of BHs and galaxy assembly.
Importantly, LRDs appear to deviate from the standard AGN paradigm in several aspects, including their 
V-shaped UV-optical spectral energy distributions (SEDs) with a turnover wavelength around rest-frame 
$\simeq 4000~{\rm \AA}$, their broad-line emission properties, and the weakness in canonical AGN 
signatures such as hot dust and X-ray corona emission. 
Moreover, LRDs host massive BHs likely heavier than the typical seed masses predicted by theory
and simultaneously have substantially undermassive host galaxies.
These features raise fundamental questions about their physical origin and role.

In the past three years since the initial discovery, a rapidly increasing number ($\sim 500$) of LRDs has been identified.
These findings have revealed several robust features, but also often delivered new and unexpected puzzles.
Given the fast pace of developments and the daily appearance of new LRD-related papers, 
key results and interpretations can sometimes be overlooked, occasionally leading to some misconceptions.
This review aims to summarize current observational and theoretical understanding of LRDs,
clarifying what aspects are well established and what remain uncertain.
We employ order-of-magnitude arguments to elucidate the governing physical processes, and
outline major open questions that define the next frontier of this emerging field.

\vspace{3mm}
\section{Key characteristics}\label{sec:character}

LRDs have drawn considerable attention due to a set of unique features that distinguish them from the population of previously known normal AGNs.
In this section, we summarize their defining observational characteristics based on current data.

%%%%%%%%%%%%%%%%%%%%%%%%%%%%%%%%%%%%%%
\vspace{3mm}
\subsection{High-redshifts objects}

We begin with the reminder that LRDs are confirmed to reside at high redshifts.
Spectroscopic observations with JWST/NIRCam and NIRSpec unveil their redshift range as $4 \lesssim z \lesssim 9$
\citep{Matthee_2024,Kokorev_2024a,Kocevski_2025}, 
corresponding to cosmic times of $t \simeq 0.5-1.5~{\rm Gyr}$ and luminosity distances of $D_{L} \simeq 30-100~{\rm Gpc}$.
Their observed fluxes, combined with these distances, imply luminosities in the rest-frame UV and optical bands on the order of
\begin{equation}
\nu L_\nu^i
=4\pi D_{L}^2~(\nu F_\nu^i)_{\rm obs}^i
\simeq 
\left\{
\begin{array}{ll}
4.0  \times 10^9~\lsun  & ~~{\rm for}~i={\rm UV}, \\
1.2\times 10^{10}~\lsun & ~~{\rm for}~i={\rm optical},\\
\end{array}
\right.
    \label{eq:nuLnu}
\end{equation}
obtained from the panchromatic SED of the typical LRD \citep{Akins_2025,Delvecchio_2025}, as shown in Figure~\ref{fig:SED}.
The typical uncertainty of the stacked flux is $8\%$ in the UV band and $6\%$ in the optical band.
This estimate neglects any dust extinction and shows the observed values without correction. 
Applying dust extinction of $A_\lambda$ at a wavelength of $\lambda$ along with an attenuation law increases 
the inferred intrinsic luminosity by a factor of $10^{0.4 A_\lambda}$ as
\begin{equation}
\nu L_\nu^i
\simeq 
\left\{
\begin{array}{ll}
6.3\times 10^{10+0.4\cdot(A_{\rm UV}-3)}~\lsun & ~~{\rm for} ~i={\rm UV}, \\
1.9\times 10^{11+0.4\cdot(A_V-3)}~\lsun & ~~{\rm for}~i={\rm optical},
\end{array}
\right.
    \label{eq:nuLnudust}
\end{equation}
where the extinction levels of $A_V$ and $A_{\rm UV}$ are defined at rest-frame ${5100~{\rm \AA}}$ and ${1500~{\rm \AA}}$, respectively.
Therefore, the optical-to-UV luminosity ratio is expressed as 
\begin{equation}
\frac{L_{\rm opt}}{L_{\rm UV}}\simeq 3.0 \times 10^{0.4\cdot(A_V-A_{\rm UV})},
    \label{eq:nuLnudust}
\end{equation}
where $L_i=\nu L_\nu^i$ ($i=$ UV and optical).

%%%%%%%%%%%%%%%%%%%%%%%%%%%%%%%%%%%%%%
\vspace{3mm}
\subsection{V-shaped SEDs}
A distinct feature of LRDs is their V-shaped UV-to-optical SEDs.
This shape is observed in the prototype LRD \citep[CEERS~746,][]{Kocevski_2023} and has become a primary selection criterion using 
JWST broad- and medium-band photometric colors \citep[e.g.,][]{Greene_2024,Kocevski_2025,Hainline_2025}.
Quantitatively, the SEDs can be described by a broken power law in the form 
$F_\lambda \propto \lambda^\beta$,
with typical slopes of $\beta_{\rm UV} \simeq -2$ (UV part) and $\beta_{\rm opt} \simeq 0$ (optical part).

This spectral characteristic motivated interpretations assuming two components in early studies on LRDs:
\begin{itemize}
    \item A blue UV slope from unobscured sources (young stellar populations or unobscured AGNs),
    \item A red optical slope from dust reddened, obscured sources (dusty galaxy or obscured AGNs).
\end{itemize}
Alternatively, a single AGN plays two effective roles:
\begin{itemize}
    \item A red optical from dust reddened, obscured AGNs and a blue UV excess from scattered light (dust and electrons) leaking through a patchy, partially covering medium.
\end{itemize}
However, such two-component models generally face a fine-tuning problem.
Despite significant variation in redshift, luminosity, and selection method, LRDs consistently exhibit remarkably similar SED shapes across samples
(see Figure~\ref{fig:SED}).
In particular, the turnover wavelength (the characteristic ``V-shaped" valley) remains narrowly confined around rest-frame 
$\lambda \simeq 3000-4000~{\rm \AA}$ \citep{Setton_2024}.
This uniformity in the SEDs is difficult to reconcile with a scenario that requires variable contributions from two unrelated components.
For instance, the explanation based on scattering processes would impose an almost universal distribution of dust and electrons in the nuclear region.
More generally, any two-component model needs to explain the physical reason for maintaining 
a tight link between the emission sources, namely early coevolution between the central BH and 
young galaxy \citep{Inayoshi_2025S}.

Moreover, dust reddening scenarios introduce a significant tension with independent infrared constraints.
Dust obscuration necessarily implies dust heating and subsequent thermal re-emission in the mid- to far-infrared \citep{Barvainis_1987}.
However, deep JWST/MIRI and ALMA observations show no significant detection of reprocessed IR flux in LRDs 
(\citealt{Williams_2024,Perez-Gonzalez_2024,Labbe_2025,Akins_2025,Delvecchio_2025}; see also the stacked SED in Figure~\ref{fig:SED}).
The absence of IR emission excess challenges any interpretation that involves heavy obscuration.

%%%%%%%%%%%%%%%%%%%%%%%%%%%%%%%%%%%%%%
\vspace{3mm}
\subsection{Compact morphology}\label{sec:morphology}

LRDs are selected as compact sources with angular sizes smaller than the point-spread function (PSF) of JWST/NIRCam imaging \citep{Barro_2024,Labbe_2025}.
Their inferred physical sizes are typically $\lesssim 100~\pc$, and the constraints reach down to $\sim 10-30~\pc$ in gravitationally lensed cases.
In most instances, LRDs show no clear signs of an extended host galaxy.
Some LRDs have no detectable extended emission consistent with stellar emission \citep{Chen_2025a},
where the tightest upper limits on stellar mass are derived based on realistic mock simulations.
A significant fraction ($\sim 30\%-50\%$) of LRDs show spatially resolved, faint components in short-wavelength filters \citep{Killi_2024,Rinaldi_2025}, 
but these are generally asymmetric and/or off-centered relative to the LRD position in longer-wavelength bands and are energetically subdominant \citep{Chen_2025a}.
Through analysis of 17 bands of NIRCam images, including a medium-band filter that captures [\ion{O}{3}]~$\lambda 5007$, at least one LRD shows extended, off-centered blue emission originating from nebular gas, not stars \citep{Chen_2025b}.
Image stacking analysis of $\gtrsim 200$ LRDs observed in the COSMOS-Web survey also reveals extended components in four NIRCam bands,
with the ratio of extended to total flux increasing toward shorter-wavelength filters (from 12\% to 40\%; \citealt{Zhang_2025_LRD_stack}).

Additionally, LRDs are often found in non-isolated environments.
High-resolution imaging frequently reveals nearby companions, either galaxies, nebular gas clumps, or even another V-shaped source 
(i.e., dual LRDs; see \citealt{Tanaka_2024b}).
These companions may serve as potential reservoirs of gas and stars that can fuel or interact with the LRDs in subsequent evolutionary phases \citep{Chen_2025b}.
The build-up of their host-galaxy components may in fact drive the disappearance of LRDs toward lower redshifts, as the initially ``naked" BHs 
lose their point-source characteristics and the characteristic V-shaped SED becomes less distinct \citep{Inayoshi_2025a,Billand_2025,Jones_2025,Merida_2025}.
Nevertheless, the majority of LRDs are not hosted by well-established, massive galaxies;
instead, their nuclear BHs appear to be overmassive relative to the local BH-to-galaxy mass
correlation, although the BH mass estimates are still highly uncertain (see Sections~\ref{sec:BHmass} and \ref{sec:MM}).

%%%%%%%%%%%%%%%%%%%%%%%%%%%%%%%%%%%%%%
\vspace{3mm}
\subsection{Variability}

Variability is a critical diagnostic for examining whether LRDs are powered by a single compact object
or by a composite of multiple components. 
AGNs commonly exhibit intrinsic flux variability, reflecting changes in accretion rates or reprocessing on short timescales \citep[e.g.,][]{Ulrich_1997,Burke_2021}.
Low-redshift highly accreting AGNs (e.g., narrow-line Seyfert 1 galaxies; NLSy1) are known to show rapid and 
low-amplitude (but nonzero) optical variability at the level of $\sim 0.1$ mag \citep{Du_2016}. 
Detecting such small variability requires high-precision flux calibration.
In contrast, galaxies composed of many unsynchronized emitters (i.e., stars) typically show minimal coherent variability unless 
multiple stellar explosions occur within the observing time window.

To date, the JWST observational programs have not been originally designed for multi-epoch observations and make the detection of variability challenging.
\citet{Kokubo_Harikane_2024} analyzed a small sample of high-redshift AGNs, including two LRDs, and found no significant variability above $\gtrsim 0.1~{\rm mag}$.
Expanding on this, \citet{Zhang_2025} analyzed $\sim 300$ publicly available LRD samples from multiple JWST surveys.
They found that the majority of LRDs do not show detectable variability (see also \citealt{Hayes_2024} and \citealt{Tee_2025} using a combination of HST 
and JWST observational data), but identified eight LRDs showing clear flux changes of $\gtrsim 0.3-0.8~{\rm mag}$ over timescales of 
$\lesssim 10-100~{\rm days}$ in the observer's frame.
This result implies that the emission originates from compact regions with physical sizes of $R\lesssim c\Delta t\lesssim 10^{-2}~\pc$,
much smaller than the PSF-limited image size of $\lesssim 100~\pc$.

One of the most intriguing LRDs for variability studies is Abell2744-QSO1, a triply gravitationally lensed source at $z=7.045$ \citep{Furtak_2024}.
Because the three lensed images correspond to path-length differences equivalent to 
a rest-frame baseline of $\sim 2.5$ years, they allow us to probe intrinsic variability at three distinct epochs within a single set of observations.
Two independent studies reported significant decreases in the equivalent widths (EWs) of H$\beta$ and H$\alpha$ emission lines \citep{Ji_2025,Furtak_2025}, although it remains unclear whether these changes were caused by continuum brightening or emission-line fading due to uncertainties in 
flux calibration and lens modeling.

Some studies have argued that the apparent lack of variability may disfavor the AGN scenario.
However, little or weak variability for LRDs is consistent with that of a subset of nearby AGN populations with stable UV fluxes
(e.g., NLSy1 galaxies).
Crucially, the detection of even a small but nonzero fraction of variable LRDs strongly supports the presence of compact, 
massive BHs at their centers.

%%%%%%%%%%%%%%%%%%%%%%%%%%%%%%%%%%%%%%
\vspace{3mm}
\subsection{Broad-line emission}\label{sec:BLR}

A key observational signature of LRDs is the presence of broad hydrogen Balmer emission lines (H$\alpha$ and H$\beta$) 
with line widths of ${\rm FWHM}\gtrsim1,000~\kms$. 
These broad-line components are considered to originate from fast-moving gas clouds in the vicinity of a central massive BH
(i.e., broad-line regions; BLRs), 
and to be a definitive characteristic of type 1 AGNs. 
Using empirical relations calibrated from nearby AGN populations, single-epoch spectroscopy of broad lines enables virial BH 
mass estimates \citep[e.g.,][]{Greene_Ho_2005,Du_2014}. 
Although these estimates are subject to uncertainties due to uncertain dust extinction (Section~\ref{sec:AGN_case})
and possible additional broadening by electron scattering (Section~\ref{sec:BHmass}), the typical BH masses lie in the range of 
$M_{\rm BH} \sim 10^{6}-10^{7}~\msun$ without extinction or scattering correction 
\citep{Matthee_2024,Maiolino_2024_JADES,Lin_2024,Taylor_2025a,Kocevski_2025}.

Moreover, JWST-identified AGNs tend to exhibit systematically higher broad H$\alpha$ EWs
\citep{Maiolino_2025}, approximately 3 times larger than those of low-redshift quasars
\citep[e.g.,][]{VandenBerk_2001,Greene_Ho_2005,Lusso_2020}.
This difference suggests that the broad-line emission mechanisms in high-redshift faint AGNs may differ from those in nearby AGNs,
for instance, through enhanced Balmer decrements in dense environments (see Section~\ref{sec:Balmer}).

It is also worth noting that the color selection criteria based on the V-shaped SEDs and compact morphologies have 
proven highly effective in identifying LRDs with broad Balmer emission lines \citep{Greene_2024}. 
A more systematic analysis by \citet{Hviding_2025} using large spectroscopic data from the RUBIES survey 
found that LRDs selected based on rest-optical compact morphologies and V-shaped SEDs are likely 
($\gtrsim 80\%$) to have broad Balmer emission lines (the remaining 20\% are not necessarily inconsistent with being broad-line AGNs).
Other studies have reported that a similarly small fraction of photometrically selected LRD candidates show clear V-shaped spectra but 
with narrow H$\alpha$ emission (${\rm FWHM}\simeq 200-300~\kms$) or with red colors dominated by strong narrow emission lines \citep{Zhang_2025_NLR}.
Despite a small number of exceptions, these findings suggest a tight physical link between the three defining features of LRDs.

Alternatively, one might attempt to explain the broad H$\alpha$ emission in LRDs through stellar phenomena,  
such as stellar winds or collections of supernova explosions (SNe).
Wolf-Rayet (WR) galaxies characterized by young and massive stellar populations can indeed produce broad H$\alpha$ 
emission \citep[e.g.,][]{Ho_1995,Schaerer_1999}.
However, emission powered by hot, massive stars (WR stars and possibly interacting binaries) would also imprint 
characteristic spectral features, frequently associated with
a \ion{He}{2}~$\lambda 4686$ bump, which are not observed in LRDs. 
Moreover, reproducing the observed H$\alpha$ luminosities of $L_{\rm H\alpha}\simeq 10^{41-42}~{\rm erg~s}^{-1}$ 
by multiple SNe would require an implausibly high rate of luminous transient events, implying extreme 
star formation rates at $\gg 100~\msunyr$ \citep[e.g.,][]{Kokubo_Harikane_2024,Maiolino_2025}.

An additional possibility of non-AGN broad lines is that ultra-dense stellar clusters could produce broad emission 
via fast-moving gas clouds with velocities of $\gtrsim 1,000~\kms$ \citep{Baggen_2024}.
However, this would require the stellar population to be confined within $\sim 100~\pc$ scales, corresponding to 
an extraordinary stellar surface density of $\Sigma_\star \simeq 10^{6-7}~\msun~{\rm pc}^{-2}$, 
orders of magnitude higher than known star-forming system (see Section~\ref{sec:stellar}).

%%%%%%%%%%%%%%%%%%%%%%%%%%%%%%%%%%%%%%
\vspace{3mm}
\subsection{Balmer absorption, break, and decrement}\label{sec:Balmer}

LRD spectra show prominent absorption features superimposed on broad-line emission of hydrogen Balmer series
and \ion{He}{1}~$\lambda 10830$ in some cases 
\citep{Matthee_2024,Maiolino_2024_JADES,Lin_2024,Juodzbalis_2024,Lin_2024,Wang_2025,Kocevski_2025,Inayoshi_Maiolino_2025,D'Eugenio_2025_abs}.
The detection of H$\alpha$ and H$\beta$ absorption is striking, as the $n=2$ excited state of atomic hydrogen is short-lived and not metastable.
To produce visible absorption against broad Balmer emission, gas densities of $n_{\rm H} \gtrsim 10^{9-11}~\cc$ are 
required to populate the $n=2$ level via collisional excitation \citep{Juodzbalis_2024,Inayoshi_Maiolino_2025}.
We note that these required gas densities are comparable to those of BLR clouds observed in nearby AGNs \citep[e.g.,][]{Kwan_Krolik_1981,Osterbrock_Ferland_2006}.
In the nearby universe, Balmer absorption is rare and seen in only $\simeq 0.1\%$ of AGNs \citep[e.g.,][]{Aoki_2006,Shi_2016,Schulze_2018}.
In contrast, JWST observations have detected Balmer absorption in at least $10\%-20\%$ of high-redshift broad-line AGNs 
\citep{Matthee_2024,Lin_2024,Kocevski_2025}.
Given that high-resolution spectroscopy is needed to identify such spectral features, the observed fraction is likely a lower limit 
\citep{Maiolino_2024_JADES}, suggesting that many AGNs are enshrouded in dense gas covering a substantial solid angle.

Several LRDs observed with the low-resolution NIRSpec PRISM mode exhibit a characteristic bump near the Balmer limit wavelength of 
$\lambda=3646~{\rm \AA}$ \citep{Greene_2024,Furtak_2024,Labbe_2024b}, 
separating the blue UV and red optical parts.
This spectral feature superficially resembles a Balmer break from evolved stellar populations and has been interpreted 
as evidence of host galaxy contributions \citep{Wang_2024b,Kokorev_2024b,Akins_2025}.
If the light blueward of the Balmer break originates from a stellar source, stellar masses could be inferred as  $M_\star \sim 10^9~\msun$, 
aligning with structure formation models (see Section~\ref{sec:stellar}).
However, some LRDs exhibit a prominent Balmer break too deep to be explained by a stellar origin \citep{Naidu_2025,deGraaff_2025b,Taylor_2025b}.
Importantly, such deep Balmer break features can be attributed to the same dense nuclear gas that produces Balmer absorption lines 
\citep{Inayoshi_Maiolino_2025,Ji_2025,D'Eugenio_2025_abs}.
This process is analogous to the Lyman break, which results from neutral atomic hydrogen in the ground state ($n=1$) absorbing ionizing radiation.

Another key feature is the elevated Balmer decrement. The observed ${\rm H}\alpha/{\rm H}\beta$ flux ratio often exceeds 
the values expected from either Case B recombination ($\simeq 2.86$; \citealt{Osterbrock_1974}) or its partially 
optically thick variants ($\simeq 3.05$; \citealt{Osterbrock_Ferland_2006}).
While such higher ratios are commonly interpreted as the result of dust reddening \citep[e.g.][]{Mathis_1970,Mathis_1983,Calzetti_1994,Dong_2008},
they may also arise from alternative processes \citep[e.g.,][]{Baldwin_1975}. 
In particular, collisional excitation and line trapping, where resonance scattering can divert H$\beta$ photons to produce Pa$\alpha$ and H$\alpha$, can enhance the Balmer line ratios without invoking significant dust attenuation
\citep{Krolik_McKee_1978,Kwan_Krolik_1981,Inayoshi_2022b,Chang_2025}. 
The latter interpretation is supported by the observations of Balmer absorption imprinted on broad emission lines in many LRDs.
These processes enhancing Balmer decrements in dense nuclear regions also align with significantly high EWs of broad H$\alpha$ emission
observed in JWST-identified AGNs (Section~\ref{sec:BLR}).
Recently, \citet{Nikopoulos_2025} extended spectral analysis to high-S/N sources by including additional Balmer lines such as H$\gamma$ and H$\delta$, and found that high Balmer decrements both in H$\alpha$/H$\beta$ and H$\alpha$/H$\gamma$ appear consistent with dust reddening with $A_V\simeq 3-5$ mag.
However, such heavy extinction would contradict the weakness of hot dust emission in their spectra (see Section~\ref{sec:NIR}).
Alternatively, dense broad-line emitting clouds with $n_{\rm H}\gtrsim 10^{10}~\cc$ can reproduce both the large Balmer decrement values and the H$\alpha$/H$\beta$-H$\alpha$/H$\gamma$ correlation.
% (Yan et al. in prep).

\begin{figure}
\centering
\includegraphics[width=180mm]{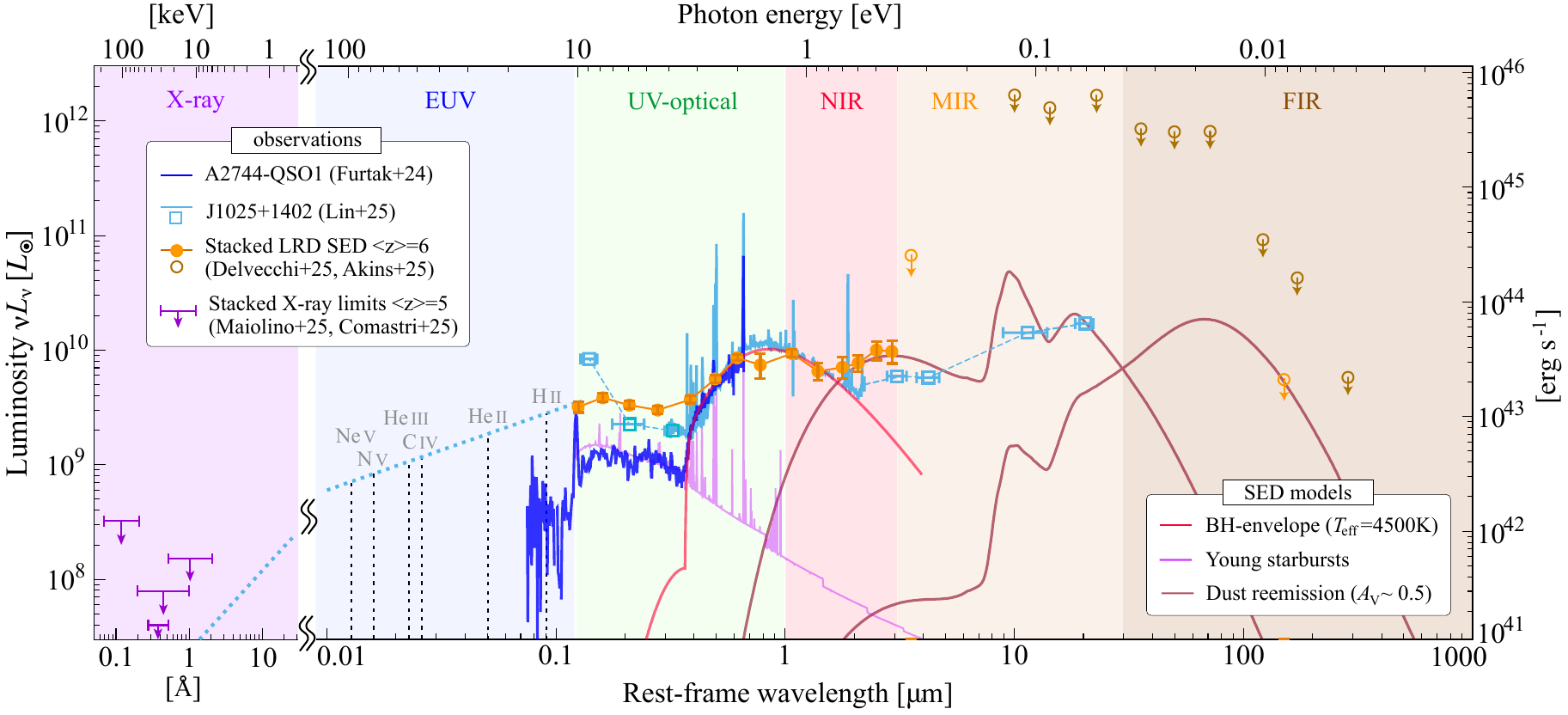}
\caption{Multi-wavelength spectra of two representative LRDs: Abell2744-QSO1 at $z=7.045$ \citep[blue,][]{Furtak_2024} and 
J1025+1402 at $z=0.1007$ with fluxes rescaled by a factor of 2.5 \citep[cyan,][]{Lin_2025c}.
Rest-frame, median-stacked panchromatic SED based on publicly available data of JWST NIRCam, MIRI, and ALMA from the 
CEERS, GOODS-S, PRIMER-COSMOS, PRIMER-UDS surveys, as well as Spitzer and Herschel upper limits 
\citep[detection/non-detection with filled/open circles,][]{Akins_2025,Delvecchio_2025}.
For comparison, the SED models of a BH-envelope with a surface temperature of $T_{\rm eff}=4500~\K$ (red), 
a young starburst (magenta), and infrared reemission from heated dust with $A_V\simeq 0.5~{\rm mag}$ (brown) are shown.
The V-shaped SED in the enshrouded-BH scenario is taken from \citet{Inayoshi_2025S} and the dust re-emission spectra
are calculated to match either the MIRI detection or the deepest ALMA non-detection limit \citep{K.Chen_2025}. 
A power-law ionizing radiation spectrum with $\alpha_{\rm ox}=-1.7$ \citep[dashed,][]{Lusso_2015} is overlaid with 
ionization energy thresholds for H, He, C, N, and Ne. 
The X-ray upper limits for JWST-identified AGNs in the Chandra Deep Field \citep{Maiolino_2025,Comastri_2025}
are shown at the shortest-wavelength end of the spectrum, where the x-axis is compressed for visualization. 
The rest-frame 5~GHz upper limit (corresponding to a wavelength of $6~{\rm cm}$) is $2\times 10^{39}~{\rm erg~s}^{-1}$
\citep{Mazzolari_2024,Gloudemans_2025}, but this value lies far below the other luminosity constraints and 
thus falls outside the plot range optimized for the other wavelength data.}
\label{fig:SED}
\end{figure}

%%%%%%%%%%%%%%%%%%%%%%%%%%%%%%%%%%%%%%
\vspace{3mm}
\subsection{Constraints from other wavelengths}

Multi-wavelength observational campaigns have been carried out to investigate the nature of LRDs, 
extending beyond the rest-frame UV-to-optical bands. 
In typical nearby AGNs, X-ray emission arises from hot plasma near the BH event horizon, near-infrared (NIR) emission 
is reprocessed by dust grains heated by AGN radiation, and radio synchrotron emission originates from relativistic jets 
and associated non-thermal processes. 
These spectral features are commonly used as diagnostics for AGNs when they are dust obscured, since high- and low-energy photons 
can penetrate dusty environments more effectively than optical and UV light \citep[e.g.,][]{Ho_2008,Hickox_Alexander2018}.
However, such canonical AGN signatures are largely absent or substantially weaker in LRDs.

\vspace{3mm}
\subsubsection{X-ray weakness}
X-ray observations are a key tool for identifying AGNs embedded in dense circumnuclear environments
\citep{Ueda_2014,Aird_2015,Nandra_2015,Ricci_2017,Ichikawa_2019} due to the transmitted nature of high-energy photons.
However, despite the presence of broad-line AGN signatures, most LRDs remain undetected even in deep Chandra X-ray data.
Stacking analyses also yield no significant signal with an upper limit of $L_{\rm X,obs} < (2-4)\times10^{41}~{\rm erg~s^{-1}}$ for JWST-identified AGNs 
\citep{Yue_2024,Ananna_2024,Maiolino_2025,Sacchi_2025,Comastri_2025}.

Hard X-rays are thought to originate from inverse Compton scattering of UV seed photons by hot nuclear plasma.
Empirically, the absorption-corrected UV and X-ray luminosities of AGNs follow a tight correlation characterized by a spectral index 
between $2500~{\rm \AA}$ (or $4.96~{\rm eV}$) and $2~{\rm keV}$, $\alpha_{\rm ox}(\equiv \D \log F_{\nu}/\D \log \nu)\simeq -1.5$ to $-1.8$
\citep[e.g.,][see the dashed line in the EUV regime of Figure~\ref{fig:SED}]{Lusso_2015}.
Assuming that the observed UV luminosity ($L_{\rm UV} \simeq 1.5 \times 10^{43}~{\rm erg~s}^{-1}$) represents the intrinsic AGN power, 
the corresponding X-ray luminosity is expected to be $L_{\rm X} \simeq (1.3-8.0)\times 10^{41}~{\rm erg~s}^{-1}$, broadly consistent 
with the current upper limits.
Conversely, if the observed UV flux reflects only a small fraction of the intrinsic emission due to dust attenuation and scattering,
the X-ray limits would suggest either intrinsic X-ray weakness ($\alpha_{\rm ox} \ll -1.8$, significantly steeper than in normal AGNs) 
or the presence of Compton-thick absorbers with $N_{\rm H}>10^{24}~{\rm cm^{-2}}$.
The hydrogen column density is 2 orders of magnitude higher than expected from dust attenuation alone.
Note that although the X-ray spectrum is heavily suppressed at $E \lesssim 20~{\rm keV}$ for such Compton-thick cases 
($N_{\rm H}\simeq10^{24}~{\rm cm^{-2}}$), the harder X-rays redshift into the observed $2-10$ keV Chandra band for high-redshift objects,
implying that even higher column densities would be required.

Intrinsic X-ray weakness is observed in nearby super-Eddington accretors \citep[e.g.,][]{Wang_2004,Dong_2012,Laurenti_2022,Tortosa_2023}, 
NLSy1 galaxies \citep{Brandt_2000,Liu_2021}, and AGNs with weak UV lines \citep{Wu_2012}.
Possible explanations include geometrically thick, super-Eddington accretion disks viewed at high inclination \citep{Pacucci_Narayan_2024}, 
efficient Compton cooling in funnel-like geometries \citep{Madau_Haardt_2024}, or cooling enhanced by mass-loaded disk winds \citep{Inayoshi_2025X}.
In any case, a tight upper limit from \citet{Sacchi_2025} suggests $\dot{M}/\dot{M}_{\rm Edd}\gtrsim3$, and the resulting soft X-ray spectra may also account for the weakness of UV emission lines in some LRDs \citep{Lambrides_2024}.

\vspace{3mm}
\subsubsection{$\Lambda$-shaped Near-infrared SEDs}\label{sec:NIR}

Infrared radiation emitted from hot dust irradiated by the LRD energy source provides key insights into their nuclear environments.
In typical AGNs, compact dusty tori surround the central accretion disks with a finite covering fraction,
reprocessing UV-optical radiation into the near- to mid-infrared (MIR) emission and producing characteristic IR excesses \citep[e.g.,][]{Honig_2010}.
If LRDs hosted similar tori, hot dust near the sublimation temperature of $T_{\rm dust}\simeq 1500~\K$ would yield bright emission 
in the JWST/MIRI bands, naturally extending their red optical continua \citep[e.g.,][]{Barvainis_1987}.
Despite extensive JWST/MIRI observations, however, most LRDs show no prominent IR excess 
\citep{Williams_2024,Perez-Gonzalez_2024,Akins_2025}, while a certain level of variations in NIR fluxes are reported among individual sources \citep{Lyu_2024}.

Deep MIRI observations provide further constraints on the LRD properties when the IR spectral shape is well captured \citep{Setton_2025b,Wang_2025}. 
In these sources, the spectrum (in $\nu L_\nu$) typically peaks at rest-frame $\lesssim 1~\mum$ and declines toward longer wavelengths with
a slope of $\beta_{\rm NIR}<-1$. 
The integrated NIR luminosity is comparable to or lower than the optical luminosity, and $\simeq 2-3$ times higher than the UV luminosity.
If the IR emission arises from reprocessed UV-optical radiation, energy conservation limits the dust extinction to 
$A_V \lesssim 1$ mag and $A_{\rm UV} \lesssim 2$ mag (\citealt{K.Chen_2025}; see the SEDs in the MIR/FIR regimes of Figure~\ref{fig:SED}). 
Steeper attenuation laws with $A_{\rm UV}/A_V> 3$ (e.g., the Small Magellanic Cloud or the Milky Way) yield even tighter limits on $A_V\ll1$ mag,
which is too low to explain the red optical continuum by dust alone.
Furthermore, the constraint on $A_V$ suggests dust-poor or dust-free nuclear environments for LRDs and limits the dust mass to 
$M_{\rm dust}\lesssim10^{4-6}~\msun$ \citep{Casey_2025,K.Chen_2025}.
Assuming a dust production efficiency per stellar mass of $f_{\rm dust,\star}\sim10^{-3}$ \citep{Schneider_Maiolino_2024}, 
the inferred low dust masses indicate only limited prior star formation, consistent with the absence of extended host galaxy features 
in JWST/NIRCam imaging (\citealt{Chen_2025a}; see Section~\ref{sec:morphology}).

If LRDs are indeed dust-poor, what produces their red optical continua?
Observationally, their optical-to-NIR spectra show ``$\Lambda$"-shaped SEDs characterized by steeply rising optical slopes and 
weak NIR emission ($\beta_{\rm opt}>0$ and $\beta_{\rm NIR}<-1$).
These spectra can be well reproduced by a single blackbody component with an effective temperature of $T_{\rm eff} \sim 5000~\K$.
While the physical origin of this thermal component remains uncertain (see Section~\ref{sec:BHE}),
similar SED shapes have been observed in low-redshift LRDs whose rest-frame NIR-MIR spectra are available (\citealt{Lin_2025c,Ji_2025b}; see its spectral and 
photometric data in Figure~\ref{fig:SED}), 
where the emission continues smoothly into the Rayleigh-Jeans tail with little hot-dust contribution.
This characteristic thermal emission implies that BHs powering LRDs may be enshrouded within an optically thick gaseous layer 
under conditions reminiscent of red giant photospheres \citep{Hayashi_1961}.

\subsubsection{Radio weakness}

Radio observations are often used to trace jet activity or other radio-emitting processes of extragalactic sources.
Deep radio data across a wide frequency range (144 MHz $-$ 10 GHz) from the Very Large Array (VLA), LOw-Frequency ARray (LOFAR), 
and MeerKAT for JWST-selected, spectroscopically confirmed broad-line AGNs reveal no detections, placing a $3\sigma$ upper limit 
at rest-frame 5 GHz of $L_{\rm 5GHz}\simeq 2\times 10^{39}~{\rm erg~s}^{-1}$ \citep{Mazzolari_2024,Gloudemans_2025}.
This upper limit remains consistent with empirical $L_{\rm X}-L_{\rm H\alpha}$ and $L_{\rm X}-L_{\rm R}$ correlations established for 
local radio-quiet AGNs \citep[e.g.,][]{Ho_1999_radio,Ho_Peng_2001,Merloni_2003,Saikia_2015}, as current radio observations are not yet 
deep enough to determine whether JWST-discovered AGNs are indeed radio weak.
In this regime, the origin of any detected radio emission would show a wide range of morphologies, spectral shapes, and brightness temperatures, 
as is commonly observed among radio-quiet AGNs \citep{Wang_An_2023}.

Deep ALMA observations have likewise failed to detect far-infrared emission expected from intense, 
dust-obscured star formation \citep{Fujimoto_2024}, providing upper limits at ALMA bands (rest-frame $100-300~\mum$ for $z\sim6$ sources)
of $\nu L_\nu<2\times 10^9~\lsun$ at ALMA 1.2~mm (e.g., \citealt{Labbe_2025,Xiao_2025,Akins_2025,Setton_2025b,Casey_2025}; see the flux upper limits at longer wavelengths in Figure~\ref{fig:SED}).
These limits place strong constraints on stellar-origin scenarios, particularly implying dust mass below $\lesssim 10^6~\msun$ 
\citep{Casey_2024,Casey_2025}, and suggest that star formation does not dominate the energetics of LRDs.

%%%%%%%%%%%%%%%%%%%%%%%%%%%%%%%%%%%%%%
\vspace{3mm}
\section{Origin of LRD power}

\begin{figure}
    \centering
    \includegraphics[width=160mm]{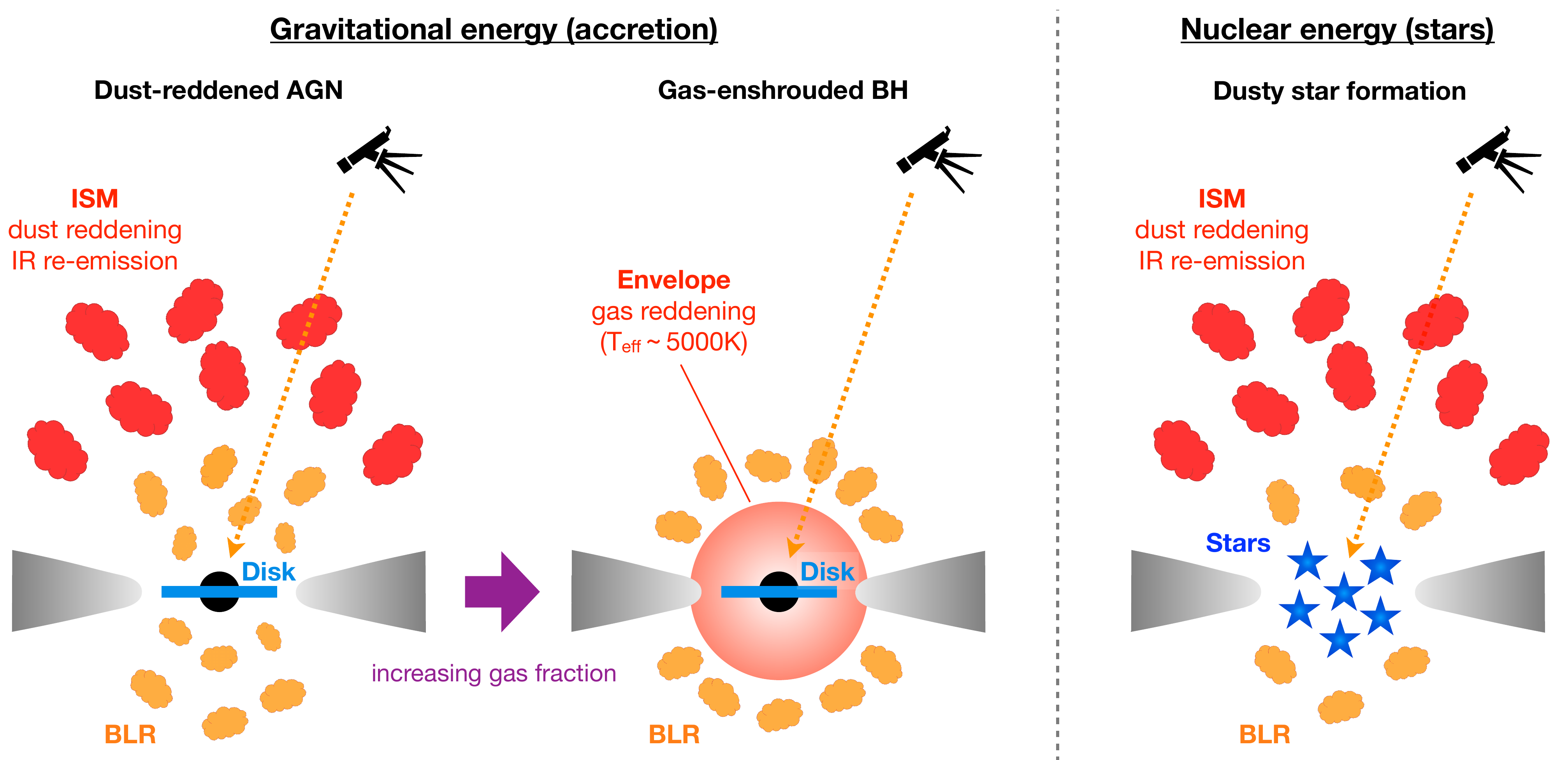}
    \caption{Summary of three proposed scenarios for the energy sources of LRDs.  
(1) \textbf{Dust-reddened AGNs}: broad-line AGNs with moderate dust extinction, explaining their red optical continua.  
(2) \textbf{Gas-enshrouded BHs}: AGNs embedded in dense gas with a high covering fraction, representing the extreme limit of an AGN + dense gas system.  
(3) \textbf{Dusty star formation}: systems where the red colors originate primarily from stellar emission attenuated by dust.  
    }
    \label{fig:scenario}
\end{figure}

Based on the observational characteristics summarized in Section~\ref{sec:character}, we investigate the physical origin of the energy source powering LRDs.
We consider two possibilities: (1) gravitational energy released through mass accretion onto a massive BH and (2) nuclear energy from 
stellar activity, recalling the debates in the 1960-1970's regarding the power source of quasars \citep[e.g.,][]{Salpeter_1964,Hoyle_1964,Fowler_1964,Sato_1966,Lynden-Bell_1969,Lynden-Bell_Rees_1971}. 
For each case, we examine whether the observed UV and optical energetics, the V-shaped SEDs, and the compact morphologies can be 
self-consistently explained without contradicting the other spectral characteristics (Sections~\ref{sec:AGN_case} and \ref{sec:stellar}, and Figure~\ref{fig:scenario}).
In short, the AGN scenario provides a more natural explanation for the observed properties, but stellar components, 
{\it if any}, may still contribute to the UV emission.

We further discuss the gas-enshrouded AGN scenario, in which an accreting BH is enshrouded within optically thick, dense gas 
with a high covering fraction, as shown in the middle panel of Figure~\ref{fig:scenario} 
\citep[e.g.,][]{Inayoshi_Maiolino_2025,Kido_2025,Liu_2025b,Naidu_2025,deGraaff_2025b,Begelman_2025,Rusakov_2025}.
This configuration can be regarded as an extreme case of the AGN + dense gas scenario, where the medium surrounding the AGN forms 
a cocoon- or envelope-like structure.
Such a configuration naturally accounts for several puzzling properties of LRDs.

%%%%%%%%%%%%%%%%%%%%%%%%%%%%%%%%%%%%%%
\vspace{3mm}
\subsection{AGN scenario}\label{sec:AGN_case}

In the AGN interpretation, the power source of LRDs is the gravitational energy released by material accreting onto a BH with mass of $M$.
The radiation luminosity can be expressed by the product of the potential energy and mass accretion rate ($\dot{M}$),
\begin{equation}
    L\simeq \frac{GM\dot{M}}{r_0} \simeq \eta\dot{M}c^2,
\end{equation}
where $r_0$ is the energy dissipation radius.
If $r_0$ is set to the inner-most stable circular orbit, the radiative efficiency becomes $\eta\simeq 0.1$ under the thin-disk approximation, 
where viscous heating is balanced with radiative cooling \citep{Shakura_Sunyaev_1973,Novikov_Thorne_1973}.

Let us begin with the optical energetics, which dominates the LRD power in the observed spectra. 
Here, we adopt a fiducial visual extinction of $A_V\simeq 3$~mag to explain the red continuum color observed in LRDs 
\citep{Kocevski_2023,Matthee_2024,Akins_2025}, assuming that the intrinsic spectrum resembles composite, low-redshift AGN SEDs 
with optical slopes of $-2\lesssim \beta_{\rm opt}\lesssim -1.5$ \citep{VandenBerk_2001,Temple_2021}.
Despite the uncertainties of $A_V$ and extinction laws, the dust-corrected optical luminosity is on the order of $L_{\rm opt}\simeq 2\times 10^{11}~\lsun$ 
(or $\simeq 7\times 10^{44}~{\rm erg~s}^{-1}$).
The mass accretion rate required to produce the optical luminosity of LRDs is 
\begin{equation}
\dot{M}=\frac{L_{\rm opt}}{\eta c^2}\simeq 0.13\eta_{0.1}~\msunyr ~10^{0.4\cdot(A_V-3)}, 
\end{equation}
where $\eta_{0.1}=\eta/0.1$.

The derived accretion rate corresponds to the Eddington accretion rate for a BH mass of $M_{\rm BH}\simeq 10^7~\msun$ at $A_V=3$.
Such inflow rates are achievable in protogalactic nuclei hosted by massive dark matter halos with masses of $M_{\rm h}$, 
where the baryonic inflow rate is given by
\begin{equation}
    \dot{M}_{\rm b}\simeq 15~\msunyr f_{\rm b,0.16}\left(\frac{M_{\rm h}}{10^{10}~\msun}\right)\left(\frac{1+z}{10}\right)^{5/2},
\end{equation}
(e.g., \citealt{Fakhouri_2010,Dekel_2013}; see more detailed derivations in \citealt{Inayoshi_2025S}), 
with the cosmic baryon fraction $f_{\rm b}(\equiv \Omega_{\rm b}/\Omega_{\rm m})=0.16,f_{\rm b,0.16}$.
Even if only a small fraction of the halo inflow reaches the nucleus (e.g., $\epsilon_{\rm nuc}\sim 0.01-0.1$; \citealt{Hopkins_Quataert_2010}), 
the resulting accretion rate ($\dot{M}=\epsilon_{\rm nuc}\dot{M}_{\rm b}$) is sufficient to power the optical energetics of 
LRDs consistently within the AGN framework.

However, explaining the UV emission of LRDs in the AGN scenario is not straightforward but requires fine-tuned conditions.
The UV continuum emission of LRDs has a blue spectral slope of $\beta_{\rm UV}\simeq -2$, 
consistent with that of unobscured quasars \citep{VandenBerk_2001}.
This similarity suggests that the continuum shape either reflects the intrinsic AGN spectrum or is affected by minimal extinction, even though 
a large $A_V$ value inferred from the red optical color leads to an inconsistent conclusion.
To reproduce such UV emission by AGN radiation alone, only a few-percent of the intrinsic UV light must escape through 
small openings from the dusty obscuring region, if LRDs are truly obscured systems
(\citealt{Kocevski_2023,Greene_2024}, as also discussed in \citealt{Zakamska_2005,Polletta_2006}).
Although this may occur in individual cases, the observed uniformity of the V-shaped SEDs across the LRD population would 
require a fine-tuned efficiency of UV-photon scattering or escape.
Otherwise, the turnover wavelength would vary significantly among sources, contrary to the observed narrow range around 
rest-frame $\lambda \simeq 4000~{\rm \AA}$ \citep{Setton_2024}.

An alternative possibility is that the UV light passes through a medium with an unusually flat extinction curve (i.e., $A_{\rm UV}\simeq A_V$), 
as observed in composite AGN spectra \citep{Gaskell_2004}, high-redshift star-forming galaxies \citep{Markov_2025}, and 
the Orion Nebula \citep{Baldwin_1991}, possibly due to a depletion of small dust grains with $\lesssim 0.01~\mum$.
In such cases, the characteristic V-shaped SED would naturally arise from the shape of the dust attenuation curve itself rather than from 
stochastic scattering or leakage effects \citep{Z.Li_2025}.

In any case, most of the UV emission (and the optical emission with $A_V\simeq 3$ mag) of AGNs is strongly attenuated by dust.
As a result, the absorbed UV-optical radiation should be reprocessed as thermal infrared emission from heated dust \citep{K.Chen_2025}.
However, JWST/MIRI observations have found faint rest-frame NIR luminosities significantly lower than the level expected 
from this scenario (Section~\ref{sec:NIR}).

%%%%%%%%%%%%%%%%%%%%%%%%%%%%%%%%%%%%%%
\vspace{3mm}
\subsection{Stellar scenario}\label{sec:stellar}

We next consider the alternative possibility that the observed emission of LRDs originates from stellar processes within compact galaxies.
As discussed below, this scenario may contribute to the UV part of the LRD spectrum, but faces problems when explaining the overall energetics and morphology.

In the stellar scenario, the observed optical luminosity can be reproduced by stars if the stellar mass is as high as 
\begin{equation}
    M_\star = \langle M_\star/L_V \rangle L_{\rm opt}\simeq 1.6\times 10^{11+0.4(A_V-3)}~\msun,
\end{equation}
where a typical mass-to-light ratio of $\langle M_\star / L_V \rangle \simeq 0.1\msun / L_{V,\odot}$ is adopted for young stellar populations 
\citep[e.g.,][]{Bell_deJong_2001, Into_Portinari_2013}, and a visual extinction of $A_V = 3$ mag is assumed to match the observed red optical continuum.
Such a high stellar mass, comparable to or exceeding that of the Milky Way ($M_{\star,\rm MW}\simeq 5\times 10^{10}~\msun$), 
has also been derived from more sophisticated SED fitting for some LRDs at redshifts $z\gtrsim 6-7$, assuming that the optical emission originates from stars 
\citep{Wang_2024b, Akins_2025}.
However, within the standard $\Lambda$CDM cosmology, such massive galaxies are expected to be exceedingly rare at these early epochs,
and even if they exist, their formation would require an extremely high star formation efficiency of $\epsilon_\star \gtrsim 30-100\%$
\citep[e.g.,][]{Inayoshi_2022c, Boylan-Kolchin_2023, Dekel_2023}.
Furthermore, the compact morphology of LRDs imposes a severe additional constraint.
The entire stellar population must be confined within $\sim 100~{\rm pc}$, yielding an extraordinarily high stellar surface density of 
$\Sigma_\star \simeq 10^{6-7}~\msun~{\rm pc}^{-2}$ \citep{Baggen_2024}. 
These densities exceed those of even the densest stellar systems known in the nearby universe (e.g., nuclear star clusters with
$\Sigma_\star \simeq 10^{5}~\msun~{\rm pc}^{-2}$; \citealt{Hopkins_2010}), and would likely lead to dynamical instabilities and gravitational 
collapse on timescales of $\lesssim 1~{\rm Myr}$ \citep{Pacucci_2025}.

The inferred stellar mass can be lowered by more than an order of magnitude if only the UV and $4000~{\rm \AA}$ break features
(excluding the red optical continua) are attributed to evolved stellar populations under negligible dust extinction.
In this case, the $4000~{\rm \AA}$ break can be explained by $\gtrsim 0.5-1.0$ Gyr-old stars (i.e., stellar Balmer break) that dominate the low-mass end of 
the initial mass function, requiring a total stellar mass of $M_\star \sim 10^{9-10}~\msun$ and a star formation rate of 
${\rm SFR} \sim 1-10~\msunyr$ \citep[e.g.,][]{Wang_2024b}. 
Although this scenario eases the tension with cosmological abundance and density limits, such systems would still lie near 
the upper limits plausible at these redshifts.

If only the UV continuum (blueward of the  $4000~{\rm \AA}$ break) originates from stellar emission, the required conditions become much less stringent.
In this case, young massive stars from ongoing star formation can reproduce the observed blue UV continuum at
\begin{equation}
    {\rm SFR}=\mathcal{K}_{\rm UV}L_{\rm UV}=1.0~\msunyr \left(\frac{\mathcal{K}_{\rm UV}}{10^{-10}~\msunyr \lsun^{-1}}\right)\left(\frac{L_{\rm UV}}{10^{10}~\lsun}\right),
\end{equation}
where the conversion factor $\mathcal{K}_{\rm UV}$ depends on the stellar initial mass function, metallicity, and stellar age
\citep{Madau_Dickinson_2014, Zackrisson_2011,Harikane_2023_galaxy}.
Adopting the fiducial conversion factor,  the UV luminosity becomes as high as $L_{\rm UV}\simeq 10^{10}~\lsun$, 
comparable to typical UV luminosities (uncorrected for dust) of observed LRDs. 
With an age of $\sim 10~{\rm Myr}$, the stellar mass of the young stellar cluster is $M_\star \sim 10^7~\msun$.
Assuming Case B radiative recombination for atomic hydrogen, the H$\alpha$ luminosity can be estimated as 
$L_{\rm H\alpha}=6.9\times 10^{41}~{\rm erg~s}^{-1}({\rm SFR}/1~\msunyr)$,
where the conversion factor is based on a $\zsun/50$ stellar population \citep{Schaerer_2002}.

\vspace{3mm}
\subsection{Enshrouded AGN scenario}\label{sec:BHE}

As discussed above, the AGN scenario is plausible in energetics but has a difficulty in explaining the IR weakness
if the optical red continuum is shaped by dust extinction.
To solve this problem, and also motivated by the presence of dense circumnuclear gas implied by 
Balmer absorption, Balmer break, and Balmer decrement, an alternative picture has been proposed in which an accreting BH is deeply 
embedded in a dense gaseous environment.
This configuration can be regarded as an extreme case of the AGN + dense-gas scenario, in which the surrounding medium forms a cocoon-like distribution (the middle panel of Figure~\ref{fig:scenario}).

In such a setup, the BH is enshrouded by an optically thick, 
extended gaseous structure. 
The obscuring medium may take the form of a quasi-hydrostatic envelope or a clumpy, geometrically 
thick gas distribution supported by the central BH engine.
Regardless of the detailed geometry and origin, such configurations naturally produce a relatively cool 
effective temperature of $T_{\rm eff} \simeq 4000-6000~\K$.
This temperature range corresponds to the minimum value attainable in an optically thick, 
partially ionized gas where H$^-$ opacity dominates.
This narrow temperature range defines the so-called Hayashi line observed as a nearly vertical track on 
the Hertzsprung-Russell diagram for cool, fully convective stellar envelopes \citep{Hayashi_1961,HHS_1962},
although this analogy applied to LRDs does not require the gas to be in a strictly stellar-like envelope.

With this interpretation, the red optical continua of LRDs can be understood as the Wien tail 
of a blackbody spectrum with $T_{\rm eff}\simeq 5000~\K$ without relying on dust reddening \citep{Inayoshi_2025S,Inayoshi_2025b,Kido_2025,Liu_2025b,Begelman_2025},
and this emission spectral shape also explains the weak IR emission 
\citep{Casey_2024,Casey_2025,Z.Li_2025, Setton_2025b, K.Chen_2025}.
Indeed, blackbody-like emission emerging from optically thick, dust-poor dense gas provides
a more natural explanation for LRDs with strong Balmer breaks and clear 
$\Lambda$-shaped SEDs in the rest-frame optical and NIR regime
\citep{Labbe_2024b,Naidu_2025,deGraaff_2025b,Taylor_2025b}.
Moreover, the dense gaseous structure accounts for the absence of X-rays due to high Compton thickness,
and plausibly also the lack of radio emission expected from AGN jets owing to effective jet energy dissipation into 
the surrounding dense medium.

Such dense gaseous structures can be supported in several physical contexts.
They may arise around BHs accreting at super-Eddington rates, where radiation pressure inflates the surrounding gas into an extended envelope \citep{Inayoshi_Haiman_Ostriker_2016,Takeo_2020,Shi_2023}.
Similar structures are also expected in quasi-stars \citep{Begelman_2006,Begelman_2008,Volonteri_2010,Coughlin_2024}
and in the brief stellar-like phase of supermassive stars formed along the direct-collapse BH pathway \citep{Hosokawa_2013,Woods_2021,Mayer_2024}.
Under the conditions where radiative energy losses become inefficient due to high optical depths,
the bloated gas resembles convection-dominated accretion flows \citep{Quataert_2000}, in which inflows 
and outflows coexist.
Convective eddies reduce the inward mass flux toward smaller radii, following a power-law form of 
$\dot{M}_{\rm in}\propto r^p$, with $p\simeq 0.5-0.7$ observed in numerical simulations
\citep[e.g.,][]{Stone_1999,Yuan_Narayan_2014}.
As a result, even if the large-scale mass inflow rate greatly exceeds the Eddington value, the BH feeding rate 
self-regulates to $\dot{M}_{\rm BH} \sim \mathcal{O}(\dot{M}_{\rm Edd})$ \citep[e.g.,][]{Hu_2022a}, and the 
emergent luminosity remains near $L_{\rm ph} \simeq L_{\rm Edd}$.

Assuming that the photospheric luminosity is close to the Eddington luminosity 
($\lambda_{\rm Edd}\equiv L_{\rm ph}/L_{\rm Edd}\simeq 1$), the BH mass is directly inferred as 
\begin{equation}
M_{\rm BH}\simeq 3\times 10^6~\msun \lambda_{\rm Edd}^{-1}\left(\frac{L_{\rm ph}}{10^{11}~\lsun}\right).
\label{eq:BHE_Mbh}
\end{equation}
The outer boundary of the obscuring gas, where the emergent continuum becomes optically thin,
defines a photospheric radius
\begin{equation}
R_{\rm ph} =\sqrt{\frac{L_{\rm ph}}{4\pi \sigma_{\rm SB}T_{\rm eff}^4}} \simeq 1.7\times 10^{16}~{\rm cm}
~\lambda_{\rm Edd}^{1/2} M_6^{1/2}T_0^{-2},
\end{equation}
where $M_6=M_{\rm BH}/(10^6~\msun)$, $T_0 =T_{\rm eff}/(5000~\K)$, and $\sigma_{\rm SB}$ 
is the Stefan-Boltzmann constant.
The circular velocity associated with the gravitational potential at this radius is
\begin{equation}
V_{\rm ph}=\sqrt{\frac{GM_{\rm BH}}{R_{\rm ph}}} 
 \simeq 3.0\times 10^{-3}c~\lambda_{\rm Edd}^{-1/4} M_6^{1/4}T_0,
 \label{eq:Vph}
\end{equation}
where $G$ is the gravitational constant and $c$ is the speed of light.
This characteristic radius is comparable to the typical BLR scale measured from 
reverberation mapping in normal AGNs \citep[e.g.,][]{Kaspi_2000,Bentz_2009}.
Gas orbiting near this radius naturally produces emission-line widths of order of $\mathcal{O}(V_{\rm ph})$.
Moderate electron scattering in the slightly deeper, optically thick layers adds an additional 
broadening component and yields the observed exponential line profiles \citep{Rusakov_2025}.

Although the BH-envelope (more generally gas-enshrouded AGNs) framework provides a self-consistent 
explanation for many key properties of LRDs, it also faces a challenge in accounting for the origin 
of the UV continuum and the broad-line emission.
A cool, optically thick gas with $T_{\rm eff}\simeq 5000~\K$ cannot generate a strong UV continuum
or produce ionizing photons.
Moreover, high-energy radiation emitted by the central accretion disk is unlikely to escape
through such dense media. 
Even if a fraction of the UV/ionizing photons emerges through low-density polar channels
or within a clumpy medium, the resulting SED would depend sensitively on the viewing angle 
and the distribution of the clumpy gas, making it difficult to maintain the spectral uniformity among LRDs.
An alternative is that the UV continuum may originate from outside the envelope, for instance, from a compact nuclear star-forming 
region surrounding the accreting BH.
While such an additional UV source can reproduce the observed UV spectra and thus the overall V-shaped SEDs, 
achieving this also requires either fine-tuning the luminosity ratio between the stellar component and 
the gas-attenuated AGN \citep{Wang_2024b,Kocevski_2025,Ma_2025a} or a physical mechanism that naturally regulates it.
Recently, \citet{Inayoshi_2025S} proposed that gas inflows during the assembly of massive halos with 
$M_{\rm h}\gtrsim 10^{11}~\msun$ at $z\sim 4-7$ trigger simultaneous BH feeding and nuclear starbursts,
whose synchronized coevolution through feedback naturally drives the luminosity and mass ratios 
between the two components to converge toward the values inferred from LRD observations.
Cumulative injection of energy and momentum by SNe subsequently expels gas from the nucleus, 
quenches gas supply to the BH envelope, and ultimately leads to a transition into normal AGN phases.

\begin{figure}
    \centering
    \includegraphics[width=180mm]{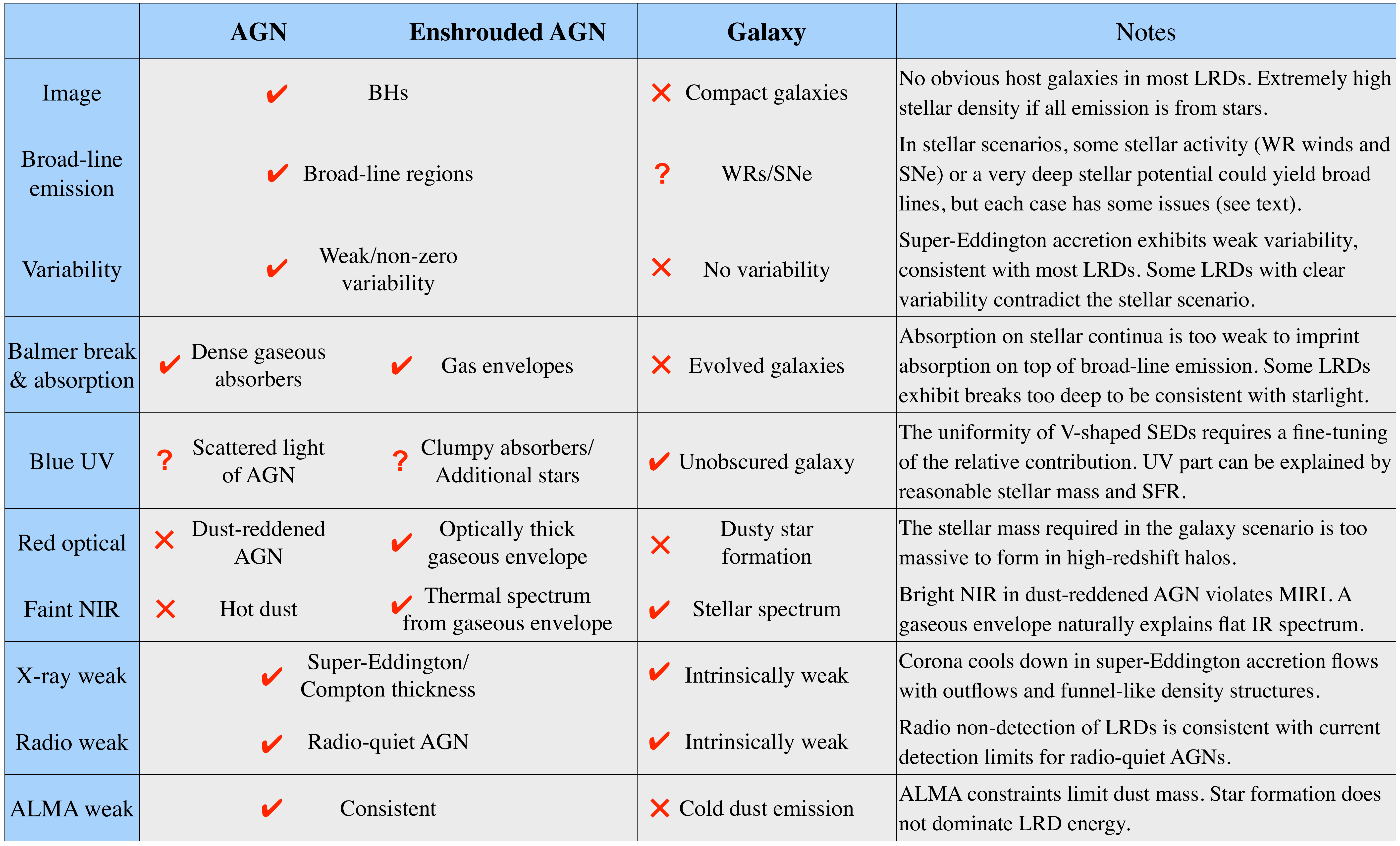}
    \vspace{0mm}
    \caption{Summary of the key observed properties of LRDs and their physical interpretations under the AGN, enshrouded AGN, and galaxy scenarios.
    		Each feature is classified as favorable ($\checkmark$), unfavorable ($\times$), or uncertain ($?$) for each model with short notes 
		explaining the rationale.
		Most observed characteristics are best explained by accreting-BH models, particularly the enshrouded-AGN framework, 
		whereas the galaxy hypothesis requires extreme fine-tuning and remains inconsistent with several constraints.
}
    \label{fig:summary}
\end{figure}

%%%%%%%%%%%%%%%%%%%%%%%%%%%%%%%%%%%%%%
\vspace{3mm}
\subsection{Summary table}

In Figure~\ref{fig:summary}, we summarize how key observational features of LRDs are interpreted under the AGN and 
galaxy hypotheses, including the gas-enshrouded AGN scenario.
Each feature is classified as favorable ($\checkmark$), unfavorable ($\times$), or uncertain ($?$) for each model.
Many observed properties, such as the compact morphology, broad Balmer emission, red optical continua,
and ALMA non-detections, are naturally explained by accreting BH scenarios.
Balmer absorption and red optical continua can arise from dense, optically thick envelopes associated 
with super-Eddington accretion, which simultaneously accounts for the weakness of X-rays and variability.
Importantly, thermal emission from such envelopes reproduces weak NIR spectra, in contrast to dust-reddened AGN interpretations.

The galaxy hypothesis, on the other hand, struggles to reproduce the lack of host signatures and the broad-line spectra without invoking 
unrealistically dense or massive stellar systems.
ALMA non-detections place a stringent limit on the SFR and dust content in LRDs.

Overall, the collective evidence favors an AGN origin, with the enshrouded AGN framework providing the most self-consistent 
explanation of the multi-wavelength properties of LRDs.

%%%%%%%%%%%%%%%%%%%%%%%%%%%%%%%%%%%%%%
\vspace{3mm}
\section{Cosmological evolution}

\subsection{Cosmic abundance}

Since their discovery, more than 500 photometrically selected LRDs have been reported, and $\sim 10\%$ of them 
have been spectroscopically confirmed as broad-line AGNs
\citep[e.g.,][]{Matthee_2024, Greene_2024,Hainline_2025, Kocevski_2025}.
The majority of the remaining sources are also likely broad-line AGNs, given that 
LRDs selected with V-shaped SEDs and compact morphologies show a high spectroscopic confirmation rate 
($\gtrsim 80\%$) for broad-line AGNs \citep{Hviding_2025}.
The large number of LRDs identified within narrow fields of view in JWST programs suggests that
their cosmic abundance reaches $\phi_{\rm LRD}\simeq 10^{-6}-10^{-4}~{\rm Mpc}^{-3}$ over $-21\lesssim M_{\rm UV} \lesssim -17$
at $z\sim 4-9$ \citep[e.g.,][]{Kokorev_2024a,Kocevski_2025,Akins_2025},
corresponding to $\sim 1\%$ of the galaxy abundance at the same brightness \citep[e.g.,][]{Finkelstein_2015,Bouwens_2021,Harikane_2022a}.

Due to the nature of their color selection, LRDs are primarily identified at $z\sim 4-9$,
where JWST multi-band filters can simultaneously capture both the blue UV and red optical continua.
The most distant confirmed LRD to date is CAPERS-LRD-z9 at $z=9.3$ \citep{Taylor_2025b}, which was
identified directly from its spectroscopic features of a V-shaped SED, a prominent Balmer break, and 
broad H$\beta$ emission with ${\rm FWHM=3,500~\kms}$, corresponding to a single-epoch BH mass of
$M_{\rm BH}\simeq 4\times 10^7~\msun$.
At even higher redshifts ($z\gtrsim 10$), characterizing the red optical continuum requires MIRI photometry,
which is generally shallower than NIRCam data and thus biases detections toward intrinsically brighter objects.
\citet{Tanaka_2025} have presented the first search for LRDs at $z\gtrsim 10$ using both NIRCam and 
MIRI F770W imaging in the COSMOS-Web field.

On the other hand, identifying LRDs at lower redshifts ($z \lesssim 3$) is also challenging.
At these redshifts, the blue UV component and the Lyman break cannot be fully captured with JWST alone,
and potential contamination from dusty red galaxies increases.
Therefore, reliable source identification requires complementary datasets from the Hubble Space Telescope and
ground-based facilities such as the Subaru Telescope.
As a result, several photometric LRD candidates have been reported at $z \sim 2-3$, yielding an estimated abundance of
$\sim 10^{-6}~{\rm Mpc}^{-3}$ \citep{Ma_2025,Euclid_LRD_2025,Zhuang_2025}, which is approximately 2 orders of magnitude lower than that at $z \sim 4-9$.
However, since none of these intermediate-redshift candidates have yet been spectroscopically confirmed as AGNs,
this abundance should be regarded as an upper limit.

The redshift evolution of the LRD abundance shows a distinct trend compared to that of normal AGNs:
the number density increases from $z\gtrsim 8$, peaks at $z\sim 6-7$, and then declines rapidly toward 
lower redshifts \citep{Kocevski_2025}.
\citet{Inayoshi_2025a} found that this evolution is well described by a log-normal distribution, 
commonly associated with phenomena driven by stochastic processes, and proposed a hypothesis that 
LRDs represent the earliest one or two accretion episodes of newly formed BH seeds.
These initial episodes exhibit characteristic spectral signatures (e.g., V-shaped SEDs) that fade 
during subsequent accretion phases.
The redshift-dependent comoving number density of LRDs is modeled in \citet{Inayoshi_2025a} as
\begin{equation}
\phi_{\rm LRD} = \phi_0 f(z) \exp\left[  -\frac{\left\{\ln (1+z)-\mu_z \right\}^2}{2\sigma_z^2}\right],
\label{eq:fit1}
\end{equation}
where $\log(\phi_0/{\rm Mpc}^{-3})=-5.27\pm0.0015$,
$\mu_z=\ln(1+z_0)$ with a characteristic redshift of $z_0~(=6.53^{+0.04}_{-0.03})$ corresponding to the cosmic age 
$t_0~(= 837\pm 6~{\rm Myr})$, and $\sigma_z =0.218\pm 0.00535$. 
The function $f(z)$ accounts for the redshift dependence of both the comoving volume and cosmic time interval,
and is approximated as
\begin{equation}
f(z)=\frac{(1+z)^{3/2}}{[s(1+z)^{1/2}-1]^2},
\label{eq:fit2}
\end{equation}
and $s=0.903$ is fitted for a flat $\Lambda$CDM universe with $\Omega_{\rm m}=0.307$ at $z\geq 1.5$.
This analytical form provides a simple model for the evolution of the LRD population,
which can be readily extended to luminosity function studies and tested with future JWST surveys, 
particularly at low and high redshifts \citep[see][]{Ma_2025}.

%%%%%%%%%%%%%%%%%%%%%%%%%%%%%%%%%%%%%%
\vspace{3mm}
\subsection{So\l tan-Paczy\'nski argument}

The high cosmic abundance of LRDs reveals that they are not rare but instead dominate the AGN population at high redshifts.
In the early stage of LRD studies, one generally interpreted LRDs as dust-reddened AGNs with moderate extinction ($A_V\sim 3~{\rm mag}$).
Using the extinction-corrected LRD luminosity function, \citet{Inayoshi_Ichikawa_2024} and \citet{Akins_2025} found that 
the BH accretion rate density does not decline but instead rises toward higher redshifts ($z\gtrsim 6$),
exceeding the value inferred from X-ray and infrared AGN surveys \citep[e.g.,][]{Ueda_2014,Delvecchio_2014,Aird_2015,Ananna_2019,Pouliasis_2024}.
This counterintuitive trend becomes even pronounced when the bright-end populations from the COSMOS-Web survey are included in the analysis \citep{Akins_2025}.

The cumulative mass density of BHs during LRD phases inferred from the radiative output with an assumed radiative efficiency
should align with the remnant BH mass density at subsequent epochs (neglecting mass loss by gravitational wave emission led by mergers).
For typical AGNs with X-ray detection at $0<z<5$, comparison with the BH mass density at $z=0$ yields a robust constraint of 
$\eta \simeq 0.1$ \citep[e.g.,][]{Soltan_1982,Yu_Tremaine_2002},
consistent with the canonical value for a geometrically thin accretion disk \citep{Shakura_Sunyaev_1973,Novikov_Thorne_1973}.
In contrast, reproducing the observed BH mass density at $z\sim4-5$ from the accretion implied by the LRD luminosity functions
requires a radiative efficiency at least twice as high ($>3\sigma$ significance), implying that these BHs rapidly spin 
up to $\gtrsim96\%$ of the maximum limit.
Improved estimates of BH mass and luminosity for LRD populations will refine the constraints on radiative efficiencies and associated spins (see Sections~\ref{sec:BHE} and \ref{sec:BHmass}).

%%%%%%%%%%%%%%%%%%%%%%%%%%%%%%%%%%%%%%
\vspace{3mm}
\subsection{Local LRD analog}\label{sec:lowz_LRD}

Although LRDs are predominantly found at high redshifts, several LRD-like objects have been reported 
in the nearby universe ($z\sim0.1-0.2$) \citep{R.Lin_2025,Lin_2025c,Ji_2025b}.
These sources exhibit hallmark LRD features: (1) a V-shaped UV-optical SED, (2) compact morphology, 
(3) broad Balmer/Paschen and \ion{He}{1} emission (${\rm FWHM}\simeq 1000-2000~\kms$) with narrow absorption, 
(4) weak NIR fluxes, (5) X-ray weakness, and (6) no significant variability.
\citet{R.Lin_2025} showed that some local ``green pea" galaxies, which are classified as compact dwarf galaxies, 
share several key properties with high-redshift LRDs, including V-shaped SEDs, broad Balmer emission, and 
BH overmassiveness relative to the local scaling relation.
For three additional, low-redshift LRD-like sources \citep{Lin_2025c,Ji_2025b}, high-resolution spectra covering 
rest-frame optical to NIR enable detailed studies of their continua and line properties, 
which are inaccessible to JWST/NIRSpec (and only marginally probed by MIRI photometry).
WISE data further constrain their MIR emission and reveal that the optical-NIR continuum emission is well reproduced by 
a blackbody with $T_{\rm eff}\simeq5000~\K$, including a Balmer break imprint consistent with radiative transfer effects \citep{Liu_2025b}.
This agrees remarkably well with the predictions from the BH-envelope model \citep{Inayoshi_2025S,Inayoshi_2025b,Kido_2025,Begelman_2025,
Nandal_Loeb_2025}.

One striking example, J1025+1402 (``The Egg"), shows extremely strong Na~D, \ion{K}{1}, \ion{Fe}{2}, and \ion{Ca}{2} triplet absorption, 
further supporting the presence of a cool, metal-enriched gas envelope.
These features as well as the optical continuum spectrum resemble those of G-to-K giant stars (even a G-band 
molecular absorption produced under extremely high gas densities of $n_{\rm H}\gtrsim 10^{14}~\cc$).
Furthermore, this object exhibits a series of optical [\ion{Fe}{2}] emission lines indicative of low-ionization, 
dense gas with $n_{\rm H}\gtrsim 10^6~\cc$ between broad and narrow-line regions \citep{Lin_2025c,Ji_2025b,Torralba_2025a}.

Such nearby LRD analogs offer unique laboratories to probe the physical structure and kinematics of the gas in the LRD interior.
Future high-resolution spectroscopic studies will be crucial to test whether the broad-line region originates from a stratified, 
multi-phase envelope and to quantify the escaping efficiency of ionizing and UV photons from these systems.

%%%%%%%%%%%%%%%%%%%%%%%%%%%%%%%%%%%%%%
\vspace{3mm}
\section{Discussion}

\subsection{BH mass estimators}\label{sec:BHmass}

The key advantage of broad-line AGNs is that their BH masses can be estimated from a single-epoch virial relation,
$M_{\rm BH}\propto ({\rm FWHM})^2 \cdot R_{\rm BLR} \propto ({\rm FWHM})^2 \cdot L^{1/2}$.
Applying this technique to LRDs with broad Balmer emission yields typical masses of $M_{\rm BH}\simeq 10^{6-8}~\msun$.
However, this calibration is based on local, unobscured broad-line AGNs where the continuum and line emissions are 
tightly correlated \citep{Greene_Ho_2005}.
Moreover, single-epoch mass estimates for super-Eddington cases would be biased toward high values \citep[e.g.,][]{Du_2014,Lupi_2024_Mbh}.
Most LRD studies have assumed moderate dust extinction ($A_V\sim 3~{\rm mag}$) inferred from the red continuum slope and,
if available, the Balmer decrement.
With extinction correction applied to the line or continuum luminosities, the inferred BH masses increase by a factor of $\simeq 10^{0.2A_V}$.
Yet, such reddening would produce bright NIR emission inconsistent with the observed JWST/MIRI data, which imply $A_V\ll1~{\rm mag}$ \citep{K.Chen_2025}.
If this is the case, the true BH masses are likely overestimated by a factor of $\sim 3-4$ in previous studies that assume $A_V\sim 3~{\rm mag}$.

Intriguingly, a significant fraction of high-S/N LRD spectra show broad-line emission profiles better described 
by an exponential rather than a single Gaussian function \citep{Rusakov_2025}.
This behavior suggests that scattering by free electrons with thermal velocity 
$v_{\rm th,e}=\sqrt{2k_{\rm B}T/m_{\rm e}}\simeq 550~\kms$ for $T=10^4~\K$ may shape the line wings \citep[see also][]{Laor_2006,Chang_2025}.
Such scattering broadens the emission lines without altering their intrinsic kinematics,
implying that the observed FWHM values are likely overestimated relative to their true widths.
As a result, BH masses derived from single-epoch virial estimators could be overestimated
by up to 2 orders of magnitude in some cases \citep{Rusakov_2025}, although a double-Gaussian fit can reproduce the line 
profiles comparably well \citep{Juodzbalis_2025}.
A caveat is that electron scattering is elastic (i.e., photon energies are conserved) and thus observed luminosities remain unaffected.
Therefore, one must still account for the observed luminosities of the broad H$\alpha$ line and 
the optical continuum emission ($\sim10^{43-44}~{\rm erg~s}^{-1}$) in the energy budget.
From this energetics requirement (see Section~\ref{sec:AGN_case}), the BH mass cannot be arbitrarily reduced but 
the number of electron scatterings is likely limited to only one to a few \citep{Rusakov_2025,Chang_2025}.
Indeed, an independent, direct dynamical BH mass measurement based on the H$\alpha$ line kinematics for Abell2744-QSO1
is consistent with the single-epoch virial estimate instead of the scattering-modified case \citep{Juodzbalis_2025_Mdyn}.

Finally, we comment on the BH mass estimate in the enshrouded-AGN model.
In this scenario, owing to the blackbody nature of the spectrum and envelope-regulated accretion,
the photospheric luminosity is expected to be on the order of $\sim \mathcal{O}(L_{\rm Edd})$, yielding a BH mass of
$M_{\rm BH}\simeq 3\times 10^{6}~\msun \lambda_{\rm Edd}^{-1}(L_{\rm ph}/10^{11}~\lsun)$ (Equation~\ref{eq:BHE_Mbh}).
Under the blackbody assumption, the bolometric correction from rest-frame {5100~\AA} is $f_{\rm bol}\simeq 1.8$, 
significantly lower than the typical AGN value ($f_{\rm bol}\simeq 9.26$; \citealt{Richards_2006}).
A similarly modest correction of $f_{\rm bol}\simeq 5.0$ is calculated by \citet{Greene_2025}, 
based on the observed multi-wavelength spectra of A2744-45924 \citep{Labbe_2024b} and RUBIES-BLAGN-1 \citep{Wang_2025}.

%%%%%%%%%%%%%%%%%%%%%%%%%%%%%%%%%%%%%%
\vspace{3mm}
\subsection{The $M_{\rm BH}$-$M_{\star}$ relation}\label{sec:MM}

In the present-day universe, a tight correlation has been observed between SMBH masses and host galaxy properties (e.g., bulge mass and stellar velocity dispersion; see reviews in \citealt{Kormendy_Ho_2013} and \citealt{Greene_ARAA_2020}), suggesting that SMBHs and galaxies have coevolved through feeding and feedback over cosmic time.
A fundamental open question is how and when this BH-galaxy relation was first established.
In this context, newly identified LRDs provide a valuable clue: their inferred BH-to-galaxy mass ratios are significantly higher than 
those in the local universe, as their host galaxies remain poorly detected, if at all, in both imaging and spectroscopy
(e.g., \citealt{Maiolino_2024_JADES,Lin_2024,Furtak_2024,Wang_2024b,Labbe_2024b,Labbe_2025,Kocevski_2025,Chen_2025a,Chen_2025b}, 
see also \citealt{Kokorev_2024b,J.Li_2025_undermassive}). 
These findings suggest that LRDs host initially overmassive BH seeds and/or experience rapid super-Eddington accretion in the early universe.

For comparison, although traditionally quasar host galaxies have also been challenging to detect under the overwhelming glare of the AGN, JWST observations have yielded better success in separating the AGN from the host, especially in low-luminosity quasars \citep[e.g.,][]{Ding_2023,Ding_2025,Stone_2024,Yue_2024_EIGER}, but also in more powerful sources with the advent of improved decomposition techniques \citep{R.Li_2025}.
Since LRDs are several orders of magnitude fainter than luminous quasars, they offer a promising opportunity to detect the underlying stellar emission.
Nevertheless, even with JWST/NIRCam, their host galaxies, if any, remain unresolved.
The inferred upper limit on their effective radius ($\lesssim100~\pc$) implies a stellar mass of $M_\star \lesssim 10^9~\msun$
\citep[e.g.,][]{Chen_2025a,Chen_2025b}.
Combining this stellar mass with single-epoch virial BH mass estimates yields $M_{\rm BH}/M_\star \gtrsim 0.01- 0.1$
\citep[e.g.,][]{Kocevski_2023,Pacucci_2023,Maiolino_2024_JADES,Lin_2024,Furtak_2024,Kocevski_2025,Jones_2025},
significantly above the local $M_{\rm BH}-M_\star$ relation \citep[e.g.,][]{Kormendy_Ho_2013,Greene_ARAA_2020,Zhuang_2023}.

The existence of overmassive BHs (or undermassive galaxies) implies that heavy BH seeds form and grow 
prior to significant host galaxy assembly \citep[e.g.,][]{Inayoshi_ARAA_2020,Volonteri_2021}.
High-resolution simulations of galactic nuclei show that overmassive BHs can naturally form via 
super-Eddington accretion in high-redshift massive dark matter halos, where dense inflows sustain accretion far above the Eddington rate \citep{Toyouchi_2021,Sassano_2023,Shi_2023,Shi_2024,Quadri_2025}.
Photon trapping and inflow ram pressure suppress radiative feedback and enable continuous BH growth 
\citep{Inayoshi_Haiman_Ostriker_2016,Park_2016,Sugimura_2017,Takeo_2020}, although mechanical feedback may intermittently limit it 
\citep{Regan_2019,Massonneau_2023,Kiyuna_2025,Lupi_2024}.
This process allows seed BHs to grow by $\sim 1-2$ orders of magnitude before substantial stellar assembly, 
yielding $M_{\rm BH}/M_\star \simeq 0.1-1$ as predicted by radiation hydrodynamic simulations
before the JWST launch \citep{Inayoshi_2022a}.

In contrast, cosmological simulations generally predict delayed BH growth, as stellar feedback prevents nuclear gas retention until galaxies 
become sufficiently massive \citep[e.g.,][]{Habouzit_2017,Angles-Alcazar_2017,Dubois_2021,Valentini_2021,Habouzit_2022,Zhu_2022}.
Limited by sub-grid feedback models, these large-scale runs tend to produce ``undermassive" BHs compared 
with the JWST-detected AGNs \citep[see][]{Habouzit_2025}, highlighting the need for improved nuclear-scale feedback treatments.
Recently, zoom-in cosmological simulations demonstrate that overmassive BHs can form via rapid 
gas accretion when the effective strengths of sub-grid feedback prescriptions are reduced \citep[e.g.,][]{Wellons_2023}.

We note that the apparent high mass ratio may partially reflect selection biases toward strong emission-line sources and systematic 
uncertainties in BH mass estimates \citep{J.Li_2025a}.
Larger and less biased samples with improved BH mass measurements will be crucial to establish the intrinsic 
BH-galaxy connection in the early universe \citep[e.g.,][]{J.Li_2025_undermassive,Ding_2025,Silverman_2025}.
Nevertheless, the presence of extremely overmassive BHs with $M_{\rm BH}/M_\star>0.1$ and potentially 
even higher in LRDs lacking host detections appears robust, as plausible combinations of systematic uncertainties (each at the $\sim 0.3$ dex level) are insufficient to bring the ratios down to the local relation.

%%%%%%%%%%%%%%%%%%%%%%%%%%%%%%%%%%%%%%
\vspace{3mm}
\subsection{Other emission lines}
\subsubsection{High-ionization lines}\label{sec:UVL}

To study the origin of the UV emission observed in LRDs and to test their spectral models,
it is useful to examine emission lines that require ionization potentials higher than the hydrogen ionization threshold
$E_{\rm H,ion}=13.6~{\rm eV}$, such as \ion{He}{2}~$\lambda 1640$, \ion{He}{2}~$\lambda 4686$,
\ion{C}{4}~$\lambda1549$, [\ion{N}{5}]~$\lambda 1240$, and [\ion{Ne}{5}]~$\lambda3426$ \citep[e.g.,][]{Nakajima_2022,Ubler_2023,Lambrides_2024,Scholtz_2025}.
For reference, the ionization potentials for several key transitions are:
\ion{He}{3} (54.42 eV), \ion{C}{4} (47.89 eV), \ion{N}{4} (47.45 eV), \ion{N}{5} (77.47 eV), \ion{Ne}{3} (40.96 eV),
\ion{Ne}{4} (63.42 eV), and \ion{Ne}{5} (97.12 eV).

So far, high-ionization emission lines have been rarely detected in JWST spectra of LRDs \citep{Labbe_2024b,Akins_2025_UV,Wang_2025b,Torralba_2025a}, 
either because the lines are intrinsically weak or the observations lack sufficient sensitivity, particularly in the faint UV regime.
Marginal detections of \ion{He}{2}~$\lambda 4686$ have been reported in some LRDs, but with very low line ratios
of \ion{He}{2}~$\lambda 4686$/H$\beta \sim 10^{-2}$, less than 5\% of the median value for local AGNs \citep{Wang_2025b}.
This observational result strongly constrains the ionizing spectral shape of LRDs.
Their spectra must include hydrogen-ionizing photons, but lack the harder radiation required to doubly ionize helium.
Such characteristics are consistent with a stellar-origin scenario for the UV emission.
In particular, young, massive stars with very low, but nonzero metallicity ($\simeq \zsun/50$) yield 
\ion{He}{2}~$\lambda 4686$/H$\beta \simeq 8\times 10^{-4}$ or a ratio between the hydrogen/ionized helium ionizing photon number fluxes,
$Q({\rm He}^+)/Q({\rm H})\simeq 5\times 10^{-4}$ \citep{Schaerer_2002}, consistent with the observed weakness 
of \ion{He}{2}~$\lambda 4686$ emission.

On the other hand, a few LRDs do show high-ionization lines.
For instance, \citet{Tang_2025} reported narrow \ion{N}{5}~$\lambda1240$ emission in an LRD at $z = 6.98$,
although neither \ion{C}{4} nor \ion{He}{2} emission lines were observed.
Because \ion{N}{5} is unlikely produced by stellar processes, its detection likely requires nonstellar ionizing photons.
One possible explanation is that the AGN radiation escapes through a funnel-like geometry aligned with the rotation axis.
However, this ``selective leakage" scenario faces a difficulty in explaining a lack of X-ray emission: 
if photons with $E_{\rm ion}>70~{\rm eV}$ can escape, even higher-energy photons ($\gtrsim$ keV X-rays) should 
also leak out since the photoionization cross section 
decreases with energy as $\sigma_\nu \propto \nu^{-3}$.

\subsubsection{Low-ionization lines}\label{sec:Low_ion}

Another useful diagnostic comes from low-ionization metal lines with $E_{\rm ion}<40~{\rm eV}$.
Classic narrow-line intensity ratio diagnostic diagrams, such as [\ion{O}{3}]/H$\beta$ vs. [\ion{N}{2}]/H$\alpha$ \citep{Baldwin_1981}
or [\ion{O}{3}]/H$\beta$ vs. [\ion{Ne}{3}]/[\ion{O}{2}] \citep{Backhaus_2022} are commonly used to classify 
galaxies as dominated by emission from AGNs or star formation. 
However, JWST-identified AGNs (including both LRDs and more typical broad-line AGNs) tend to lie along the loci 
occupied by star-forming galaxies on these diagrams \citep{Kocevski_2023,Maiolino_2024_JADES,Scholtz_2025}.
This behavior is broadly consistent with photoionization models for low-metallicity AGNs, where reduced metal-line cooling and 
altered relative element abundances (e.g., N/O $\propto Z$) shift line ratios toward the star-forming region \citep{Hirschmann_2019}.

Nevertheless, some LRDs show intrinsically weak [\ion{O}{3}] emission.
For instance, Abell2744-QSO1 has a clear detection of [\ion{O}{3}]~$\lambda 5007$ in its high-resolution spectrum \citep{Ji_2025}.
However, the [\ion{O}{3}]/H$\beta$ ratio is extremely low, indicative of a very low gas-phase metallicity of $\sim 0.01~\zsun$ \citep{Maiolino_2025_lowZ}.
The weakness of [\ion{O}{3}] emission is attributed to spatially extended, less dense regions where collisional de-excitation would unlikely suppress the [\ion{O}{3}] line,
suggesting its metal-poor environment.

Complementing the oxygen lines, most LRDs also lack \ion{Fe}{2} emission from both permitted and forbidden transitions
(\citealt{Trefoloni_2025}; but see also strong \ion{Fe}{2} emission in some LRDs; \citealt{Labbe_2024b} and \citealt{Torralba_2025b}), 
even though \ion{Fe}{2} emission is normally prominent in highly accreting, low-mass AGNs
\citep{Greene_Ho_2007,Shen_Ho_2014}.
The deficit of iron and $\alpha$-elements (e.g., C, O) thus reflects incomplete chemical enrichment in LRD nuclei,
compared to quasar populations \citep[e.g.,][]{Nagao_2006,Onoue_2020,Jiang_2024_MgFe}.
Furthermore, when AGNs are embedded in dense gas, 
resonance fluorescence of Ly$\beta$ ($n=3\rightarrow 1$) excites neutral oxygen and produces the \ion{O}{1} 
($\lambda 1304$, $\lambda 8446$, and $\lambda 11287$) emission lines \citep{Kwan_Krolik_1981}.
These \ion{O}{1} lines are observed in low-redshift AGNs as tracers of dense circumnuclear regions and correlate well with other low-ionization 
features such as \ion{Ca}{2} and \ion{Fe}{2} \citep[e.g.,][]{Grandi_1980,Rodriguez-Ardila_2002a,Riffel_2006,Matsuoka_2007,Matsuoka_2008,Martinez-Aldama_2015,Cracco_2016}.
So far, two of the \ion{O}{1} emission lines ($\lambda 8446$ and $\lambda 11287$) have been detected in some JWST-identified AGNs: 
GN-28074 at $z = 2.26$ \citep{Juodzbalis_2024} and RUBIES-BLAGN-1 at $z = 3.1$ \citep{Wang_2025}.
Recently, \citet{Tripodi_2025} reported an LRD at $z=5.3$ with the first high-redshift detection of
the \ion{O}{1}~$\lambda1304$ bump (i.e., blend of \ion{O}{1}~$\lambda 1304$ and \ion{Si}{2}~$\lambda 1263$) together with \ion{O}{1}~$\lambda 8446$ emission.
Radiation-hydrodynamic simulations have suggested that detection of these \ion{O}{1} lines with JWST/NIRSpec would provide a direct probe of
dense, super-Eddington accretion flows around seed BHs \citep{Inayoshi_2022b}.

%%%%%%%%%%%%%%%%%%%%%%%%%%%%%%%%%%%%%%
\vspace{3mm}
\subsection{Environments}

A useful approach to understanding the formation of LRDs and their connection to quasars in a cosmological context is to determine
how these systems emerge within the evolving dark matter cosmic web.
The clustering strength of a given population, or equivalently the size of the cosmic overdensities that they inhabit, 
provides a direct estimate of their typical host halo masses.
Imaging analyses indicate that LRDs are not isolated but instead reside in moderately clustered environments, in contrast to 
UV-bright quasars whose host halos are expected to reach $M_{\rm h}\sim10^{12}$–$10^{13}~\msun$ \citep[e.g.,][]{Pizzati_2025}.
Clustering analyses for LRDs with typical luminosities suggest substantially lower halo masses of 
$M_{\rm h}\simeq (0.6-3)\times10^{11}~\msun$ at $z \simeq 4-6$, consistent across multiple independent studies
\citep{Arita_2025,X.Lin_2025_cluster}.
Notably, \citet{Schindler_2025} recently reported the brightest LRD at $z=7.26$ ($L_{\rm bol}\simeq 4\times 10^{45}~{\rm erg~s}^{-1}$, 10 times higher if correcting dust reddening with $A_V\sim 2.8$ mag) embedded in a pronounced overdensity comparable to
that of UV-luminous quasars at $z\sim6$, implying a host halo mass of $M_{\rm h}\simeq10^{12}~\msun$, the most strongly 
clustered LRD identified to date.

From the dark matter halo mass function \citep{Sheth_Tormen_2002}, the number density of such halos exceeds the observed LRD abundance
($\phi_{\rm LRD}\simeq 6\times10^{-5}~{\rm Mpc^{-3}}$ at $z\sim 4-8$; \citealt{Kocevski_2025,Kokorev_2024a,Zhuang_2025}) 
by 2 orders of magnitude, implying a duty cycle of only $\sim1\%$.
This value is comparable to that of UV-luminous quasars \citep[e.g.,][]{Davies_2019,Eilers_2021,Eilers_2024}, suggesting that the LRD phase represents 
a short-lived stage of BH growth.
Such brief active phases challenge current models of early BH formation, indicating that most BH mass assembly may occur in highly obscured 
or radiatively inefficient accretion episodes preceding the visible LRD stage.

%%%%%%%%%%%%%%%%%%%%%%%%%%%%%%%%%%%%%%
\vspace{3mm}
\subsection{Alternatives}

The lack of a definitive conclusion on the origin of early BHs leaves room for alternative, non-baryonic seeding scenarios.
In the absence of clear signatures of host galaxies, gravothermal collapse in self-interacting dark matter (SIDM) halos provides 
a natural pathway for forming BHs in high-concentration halos with a characteristic mass scale determined by the SIDM cross section
\citep{Balberg_2002,Feng_2021_SIDM,Grant_Roberts_2025}.
This process occurs preferentially in the early universe and could account for the observed abundance of LRDs, apparently galaxy-less, 
early AGN hosting massive BHs \citep{Jiang_2025_SIDM}.
Another possibility involves primordial black holes (PBHs), which can form from the collapse of early-universe density fluctuations or 
topological defects such as cosmic strings \citep[e.g.,][]{Carr_1975}, or during phase transitions that temporarily render the universe pressureless
(i.e., matter-dominated).
For PBHs formed via critical collapse, the resulting mass spectrum depends sensitively on the primordial fluctuation power spectrum.
A nearly monochromatic spectrum \citep[e.g.,][]{Yokoyama_1998} produces PBHs with a narrow mass range, insufficient to explain both dark 
matter and supermassive BHs in galactic nuclei, whereas scale-invariant fluctuations yield a broader, power-law mass distribution \citep{Harada_2016}.
PBH-seeded models would result in extremely high $M_{\rm BH}/M_\star$ values and describe low-metallicity environments of some LRDs \citep{Dayal_2025,Maiolino_2025_lowZ}

%%%%%%%%%%%%%%%%%%%%%%%%%%%%%%%%%%%%%%
\vspace{3mm}
\section{Summary}

LRDs are a newly discovered AGN population revealed by deep JWST observations.
Their properties, such as compact morphology, red optical continua, and V-shaped SEDs,
differ from the canonical AGN framework and have opened new perspectives on the 
earliest phases of BH growth.
This review summarizes recent progress in understanding how accreting BHs can reproduce 
the characteristic LRD features,
how their nuclear environments differ from those of normal AGNs, and how forthcoming observations
can discriminate among competing scenarios.

JWST spectroscopic observations show that LRDs exhibit broad Balmer emission lines with ${\rm FWHM} \gtrsim 1,000~\kms$, 
consistent with the presence of accreting BHs with $M_{\rm BH}\simeq10^{6}-10^{7}~\msun$.
In most cases, the optical luminosity of LRDs appears difficult to explain with stellar emission alone, although young stellar 
populations may still contribute to the UV continuum.
The coexistence of broad emission with Balmer absorption and prominent Balmer breaks implies that the nuclear regions
are heavily embedded in a dense gaseous medium with a high covering fraction.
One promising interpretation is that LRDs host gas-enshrouded AGNs.
In this scenario, a dense, potentially clumpy, gaseous envelope around the nucleus provides both gas attenuation 
and thermal self-emission with an effective temperature of $T_{\rm eff}\simeq 5000~\K$.
The resulting continuum emission accounts for the red optical and flat NIR spectra without requiring substantial dust reddening.
Such a configuration also explains Balmer absorption and weak (short-term) variability, offering a possible connection to 
super-Eddington accretion episodes.

From both their spectral properties and their cosmic abundance evolution (rising from $z>10$ toward $z\sim6-7$ and declining to lower redshifts),
LRDs can be interpreted as a transient phase in early BH growth, possibly corresponding to the first one or two accretion episodes of newborn BHs.
As such, they may represent the long-sought missing link between BH seed formation and the emergence of luminous quasars.

Future observations will be crucial to test these scenarios.
Improved monitoring campaigns to probe long-term variability, IFU observations to resolve nuclear 
gas kinematics and to map the spatial distribution of diagnostic emission lines, and clustering analysis of LRDs will provide key diagnostics 
of the central engine and its environment.
In addition, polarization measurements, searches for post-LRD descendants, and 
studies of local analogs will offer complementary constraints on their physical nature.
High-resolution spectroscopy and multi-wavelength campaigns with JWST, ALMA, and next-generation facilities 
will further constrain their energetics, metallicity, and feedback signatures.
Finally, joint observations with future gravitational-wave experiments may probe the emergence of the earliest BHs, 
including channels involving non-baryonic origins.

%% Please use the acknowledgment and contribution environments. This will 
%% be anonomyized when the "anonymous" style option is used. 
\begin{acknowledgments}

We greatly thank Chang-Hao Chen, Subo Dong, Sihan Lu, Hao Wu, and Zijian Zhang for constructive discussions.
K.I. and L.C.H. acknowledge support from the National Natural Science Foundation of China (12573015, 1251101148, 12233001, 12473037), and the China Manned Space Program (CMS-CSST-2025-A09).

\end{acknowledgments}

\begin{contribution}
%%This section gives authors the space to recognize author contributions. The text inside this environment is NOT counted towards the total word quanta. At a minimum, manuscripts are expected to include this text:

Both authors contributed equally.

\end{contribution}

%\appendix

%% For this sample we use BibTeX plus aasjournalv7.bst to generate the

\bibliography{ref}{}
\bibliographystyle{aasjournal}



\end{document}